\UseRawInputEncoding
\documentclass[prb]{revtex4-2}
\usepackage{threeparttable}
\usepackage{booktabs}
\usepackage{float}
\usepackage{multirow}
\usepackage{graphicx}% Include figure files
\usepackage{dcolumn}% Align table columns on decimal point
\usepackage{bm}% bold math
\usepackage{hyperref}% add hypertext capabilities
\usepackage{bookmark}
\usepackage{color}
\definecolor{red}{rgb}{0.8, 0.0, 0.0}

\begin{document}

% \preprint{APS/123-QED}

\title{General-Purpose Machine-Learned Potential for CrCoNi Alloys Enabling Large-Scale Atomistic Simulations with First-Principles Accuracy}

\author{Yong-Chao Wu}
\email{Contact author: yongchao.wu@aalto.fi}
\author{Tero M\"{a}kinen}
\author{Mikko Alava}
\affiliation{Aalto University, Department of Applied Physics, PO Box 11000, 00076 Aalto, Espoo, Finland}

\author{Amin Esfandiarpour}
\email{Contact author: amin.esfandiarpour@ncbj.gov.pl}
\affiliation{
 NOMATEN Centre of Excellence, National Center for Nuclear Research, A. Soltana 7 St., 05-400 Otwock, Poland
}

\date{\today}

\begin{abstract}
CrCoNi medium-entropy alloys exhibit exceptional mechanical properties arising from pronounced chemical complexity, including short-range order (SRO), and low stacking fault energy, posing challenges for large-scale atomistic simulations. While most models focus on equimolar compositions, deviations from equimolarity provide an effective route to tuning properties, requiring transferable interatomic potentials that capture composition-dependent behavior. Here we develop a general-purpose machine-learned interatomic potential for the CrCoNi system within the neuroevolution potential (NEP) framework, achieving near first-principles accuracy with high computational efficiency. Trained on a comprehensive dataset spanning pure elements, binary and ternary alloys across a wide compositional range, diverse crystal structures and thermodynamic conditions, and based on spin-polarized \textit{ab initio} data, the model accurately reproduces equations of state, phonons, elastic constants, dislocation dissociation, surface and defect energies, melting temperatures and strain-induced phase transformations. It further captures SRO and its effect on stacking fault energies across both equimolar and non-equimolar compositions, in agreement with first-principles and experiments. In contrast to existing potentials, typically limited to equimolar alloys and less accurate for pure elements, the present model delivers consistent accuracy across the full compositional space while retaining superior efficiency. These results enable reliable atomistic simulations of composition-dependent behaviour and provide a framework for the design of non-equimolar CrCoNi alloys.
\end{abstract}

\maketitle

\section{Introduction}

High- and medium-entropy alloys (H/MEAs) have attracted significant attention in recent years due to their vast compositional design space and their potential to exhibit exceptional mechanical and physical properties~\cite{george2019high, zhang2024frontiers, esfandiarpour2025ML}. Among these systems, the CrCoNi alloy has emerged as a model material because of its outstanding mechanical performance, including exceptional strength, ductility, and fracture toughness over a wide temperature range, from cryogenic to ambient conditions~\cite{li2019mechanical, liu2022exceptional}. In addition, experiments involving dynamic impact, laser shock, and irradiation have demonstrated its remarkable resistance to damage, impact loading, and radiation-induced defects~\cite{yang2019highimpact, zhao2023deformation, granberg2016mechanism}. 
These exceptional properties originate from the interplay of multiple deformation mechanisms, including dislocation slip, deformation twinning, and strain-induced phase transformation from face-centered cubic (FCC) to hexagonal close-packed (HCP) structures~\cite{liu2022exceptional, yang2019highimpact, slone2018influence}. The activation of these mechanisms is highly sensitive to temperature and the local chemical environment, which influence the relative stability of competing phases~\cite{coury2021multi}. In particular, chemical short-range order (SRO) has been identified as a key factor governing stacking fault energy (SFE), dislocation behavior, and phase stability, thereby strongly affecting the overall mechanical response~\cite{zhu2023effects, zhang2020short}.

Beyond equiatomic compositions, recent studies have demonstrated that deviations from equimolar or near-equimolar ratios provide an effective strategy for further enhancing mechanical properties. In particular, non-equimolar CrCoNi alloys have been shown to achieve significantly higher strength while maintaining excellent ductility. For example, Coury \textit{et al.} reported that a Cr$_{45}$Co$_{27.5}$Ni$_{27.5}$ alloy exhibits a yield strength more than 50\% higher than that of equimolar CrCoNi, with comparable ductility~\cite{coury2018high}. Similar observations have been reported in subsequent studies, highlighting the critical role of composition in tuning phase stability, SFE, and deformation mechanisms~\cite{yan2023design}. These findings indicate that the vast compositional space of CrCoNi alloys offers substantial opportunities for materials design beyond the equimolar limit. However, understanding and predicting composition-dependent behavior at the atomistic scale remain major challenges. Despite extensive efforts, the fundamental atomistic origins of deformation and strengthening mechanisms in CrCoNi alloys remain incompletely understood. A notable discrepancy persists between theoretical predictions and experimental measurements of SFE: density functional theory (DFT) calculations typically predict negative values, whereas experiments report positive values~\cite{zhang2020short}. Whether this inconsistency arises from the presence of SRO~\cite{zhang2020short}, strong solute--dislocation interactions~\cite{shih2021stacking}, or other factors remains under active debate. Moreover, the role of SRO in strengthening, whether it enhances~\cite{zhang2020short} or has negligible effects~\cite{rasooli2026searching, li2023evolution}, has not yet reached a consensus. Large-scale atomistic simulations provide a powerful approach for probing the nanoscale chemical environment and revealing the atomic origins of the exceptional mechanical properties. Although quantum-mechanical methods such as DFT offer high accuracy, their application is limited to relatively small system sizes and short time scales. In contrast, classical molecular dynamics (MD) and Monte Carlo (MC) methods are significantly more computationally efficient but rely critically on the accuracy of interatomic potentials. For metallic systems, the embedded-atom method (EAM)~\cite{daw1984embedded} and the modified embedded-atom method (MEAM)~\cite{lee2000secondn} have been widely used over the past decades, particularly for elemental metals and their alloys. However, these empirical potentials often lack the accuracy required to reliably describe complex chemical interactions due to their constrained functional forms. 

Recently, machine-learning interatomic potentials (MLPs)~\cite{mishin2021machine, du2022chemical} have emerged as a promising alternative. In MLPs, the potential energy surface is represented using flexible machine-learning models, enabling the use of a substantially larger number of fitting parameters compared with traditional many-body potentials. Freed from rigid analytical functional forms, MLPs can achieve significantly improved accuracy and transferability across diverse atomic configurations. A machine-learning potential based on the Moment Tensor Potential (MTP)~\cite{novikov2022magnetic, evgeny2023mlip3} framework has recently been developed for the CoCrNi system~\cite{cao2025capturing}. While this model demonstrates good performance for equimolar ternary alloys, its applicability remains largely restricted to near-equimolar compositions and does not provide reliable accuracy for the constituent unary elements or for compositionally off-equimolar alloys. This limitation hinders its use in predictive simulations aimed at exploring composition-dependent properties and designing non-equimolar alloys. In addition, the computational cost of MLP is still significantly higher than that of traditional empirical potentials, which can limit its application in large-scale simulations. To the best of our knowledge, a transferable and computationally efficient MLP capable of accurately describing both unary elements and the full CrCoNi compositional space, including non-equimolar alloys, is still lacking.

In this work, we develop a general-purpose machine-learning potential for the CrCoNi system based on the neuroevolution potential (NEP)~\cite{fan2021neuroevolution, song2024general} framework. The proposed model is trained on a diverse dataset spanning unary, binary, and ternary configurations, enabling accurate reproduction of key structural, thermodynamic, and mechanical properties across the compositional space. Importantly, the dataset explicitly includes non-equimolar compositions, ensuring transferability beyond the equiatomic limit. Comprehensive benchmarking demonstrates that the present model outperforms widely used empirical potentials and existing MLPs while maintaining high computational efficiency suitable for large-scale atomistic simulations. This enables predictive simulations of composition-dependent properties across the full CrCoNi compositional space.

\section{Results}

In this section, we evaluate the performance of the NEP model with respect to several key properties. For comparison, we consider several representative classical interatomic potentials, including two EAM potentials developed by Li \textit{et al.}~\cite{li2019strengthening} (denoted as EAM1) and Farkas \textit{et al.}~\cite{farkas2020model} (denoted as EAM2), as well as a MEAM potential developed by Choi \textit{et al.}~\cite{choi2018understanding}. These empirical potentials have been widely used in previous studies to investigate the mechanical properties of CrCoNi alloys~\cite{han2024ubiquitous, hua2023revealing, yan2023effectof, zhu2021unprecedented, utt2022origin, amin2022edge}. In addition, a recently developed MTP model (TS-f version, as recommended) by Cao \textit{et al.}~\cite{cao2025capturing} is included as a representative example of modern machine-learning interatomic potentials.

\subsection{Energy, force and stress validation}

\begin{figure}[H]
    \centering
    \includegraphics[width=0.85\linewidth]{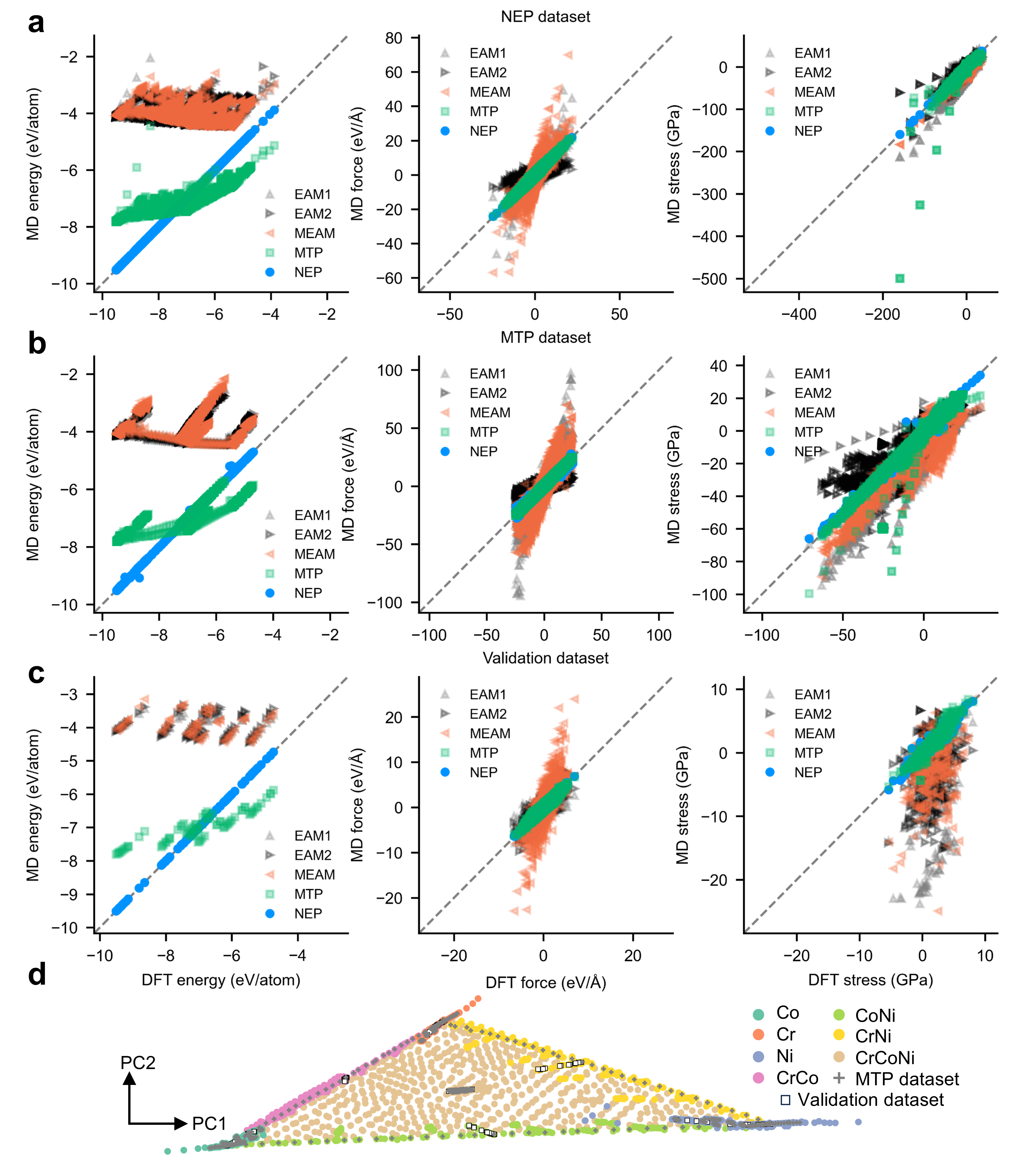}
    \caption{Pairwise comparisons of energies, forces, and stresses for the (a) NEP training dataset, (b) MTP dataset~\cite{cao2025capturing}, and (c) validation dataset, as predicted by NEP, EAM1~\cite{li2019strengthening}, EAM2~\cite{farkas2020model}, MEAM~\cite{choi2018understanding}, and MTP~\cite{cao2025capturing}. (d) Principal component analysis (PCA) of the dataset based on NEP descriptors, where each point represents an individual atomic configuration.}
    \label{fig:dataset}
\end{figure}

We first evaluate the predictive performance of these potentials for energies, forces, and stresses using three datasets: the NEP training dataset, the MTP dataset, and an independently prepared validation dataset. The NEP dataset contains 3030 diverse atomic configurations, while the MTP dataset contains 5798 structures. The validation dataset was generated from 2 ns MD simulations conducted over a temperature range from 5 to 3000 K using seven different initial structures. Detailed procedures for dataset generation are provided in the Methods section. As shown in Fig.~\ref{fig:dataset}, NEP exhibits the best overall agreement with DFT results for energies, forces, and stresses across all three datasets, indicating excellent transferability and high accuracy. MTP also demonstrates good accuracy in force prediction but shows relatively poor performance in describing stresses. Interestingly, MTP correctly predicts the energies for only a portion of the dataset, primarily corresponding to CrCoNi ternary alloy structures, while exhibiting large discrepancies under other conditions. In contrast, all empirical potentials show significant deviations from the DFT results. The root mean square errors (RMSEs) of the various potentials are summarized in Table~\ref{tab:error}. For force prediction, NEP achieves the lowest RMSE values across all datasets: 106.1 meV/\AA\ on the NEP dataset, 144.04 meV/\AA\ on the MTP dataset, and 121.41 meV/\AA\ on the validation dataset. MTP exhibits a comparable force prediction performance on the MTP dataset (144.77 meV/\AA), but shows larger RMSE values on the NEP dataset (153.76 meV/\AA) and the validation dataset (192.41 meV/\AA) compared with NEP. Among the empirical potentials, EAM1 provides the best force prediction performance, followed by EAM2 and MEAM across all datasets. 

To further analyze the structural diversity of the datasets, principal component analysis (PCA) was applied to reduce the dimensionality of the NEP descriptor space for the three datasets. The first two principal components account for 97\% of the explained variance, indicating that they capture the dominant structural features of the datasets. As illustrated in Fig.~\ref{fig:dataset}d, the NEP dataset exhibits a broader and more uniform distribution in descriptor space, indicating greater structural diversity. In contrast, the MTP dataset is mainly concentrated near the corners and central regions, lacking configurations corresponding to non-equiatomic ternary alloys, despite having a larger total number of structures than the NEP dataset.

\begin{table*}
\centering
\caption{Root Mean Square Error (RMSE) of energy (meV/atom), force (meV/\AA), and stress (GPa) predicted by EAM1~\cite{li2019strengthening}, EAM2~\cite{farkas2020model}, MEAM~\cite{choi2018understanding}, MTP~\cite{cao2025capturing}, and NEP on the NEP, MTP, and validation datasets.}
\label{tab:error}
\small
\setlength{\tabcolsep}{6pt} % optional fine control
\begin{tabular*}{\textwidth}{l@{\extracolsep{\fill}}ccccccccc}
\toprule
Potential & \multicolumn{3}{c}{NEP Dataset} & \multicolumn{3}{c}{MTP Dataset} & \multicolumn{3}{c}{Validation Dataset} \\
\cmidrule(lr){2-4} \cmidrule(lr){5-7} \cmidrule(lr){8-10}
& Energy & Force & Stress & Energy & Force & Stress & Energy & Force & Stress \\
\midrule
EAM1  & 3154.27 & 419.86 & 9.55 & 3087.76 & 699.59 & 7.63 & 3283.37 & 453.48 & 7.91 \\
EAM2  & 3134.91 & 728.65 & 5.82 & 3142.17 & 1085.47 & 5.26 & 3316.21 & 829.59 & 4.88 \\
MEAM  & 3192.76 & 800.29 & 4.41 & 3189.40 & 1125.57 & 5.03 & 3356.37 & 988.00 & 5.06 \\
MTP   & 775.41  & 153.76 & 5.74 & 547.57  & 144.77  & 2.39 & 866.23  & 192.41 & 0.72 \\
NEP   & 2.45    & 106.10 & 0.39 & 10.65   & 144.04  & 0.64 & 3.10    & 121.41 & 0.42 \\
\bottomrule
\end{tabular*}
\end{table*}

\subsection{Basic crystal properties}

\begin{figure}
    \centering
    \includegraphics[width=0.65\linewidth]{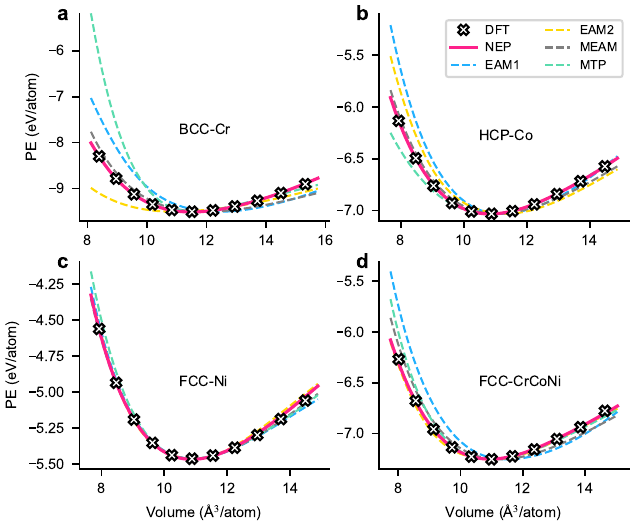}
    \caption{Energy–atomic volume relations for (a) BCC Cr, (b) HCP Co, (c) FCC Ni, and (d) CrCoNi alloy, as obtained from DFT, NEP, EAM1~\cite{li2019strengthening}, EAM2~\cite{farkas2020model}, MEAM~\cite{choi2018understanding}, and MTP~\cite{cao2025capturing} calculations.}
    \label{fig:eos}
\end{figure}

We next evaluate the fundamental physical properties of the elemental metals and the ternary CrCoNi alloy to further assess the accuracy of the developed NEP model. Fig.~\ref{fig:eos} shows the energy--volume (equation of state, EOS) curves for body centered cubic (BCC) Cr, HCP Co, FCC Ni, and equiatomic CrCoNi MEA obtained from interatomic potentials and DFT calculations. For the elemental metals, $2 \times 2 \times 2$ primitive supercells are used, while a $3 \times 3 \times 3$ conventional supercell is employed for the CrCoNi alloy. The energies and volumes are reported per atom, and identical atomic structures are used in both the potential-based calculations and the DFT reference to ensure a consistent comparison. Except for NEP, the energy curves of all other potentials have been shifted so that their minimum energies coincide with the DFT minimum. The NEP predictions show excellent agreement with DFT across the entire volume range examined, corresponding to volumetric strains of $\pm 10\%$. Accurate EOS curves are essential for describing the mechanical response of materials under a wide range of volumetric strains and pressures. While all interatomic potentials reproduce the EOS of Ni reasonably well and show relatively small deviations for CrCoNi, noticeable discrepancies are observed for Cr and Co when using empirical potentials and MTP, particularly under compression. These results indicate the limited capability of these potentials in accurately describing the EOS of the elemental metals.

\begin{figure}
    \centering
    \includegraphics[width=0.9\linewidth]{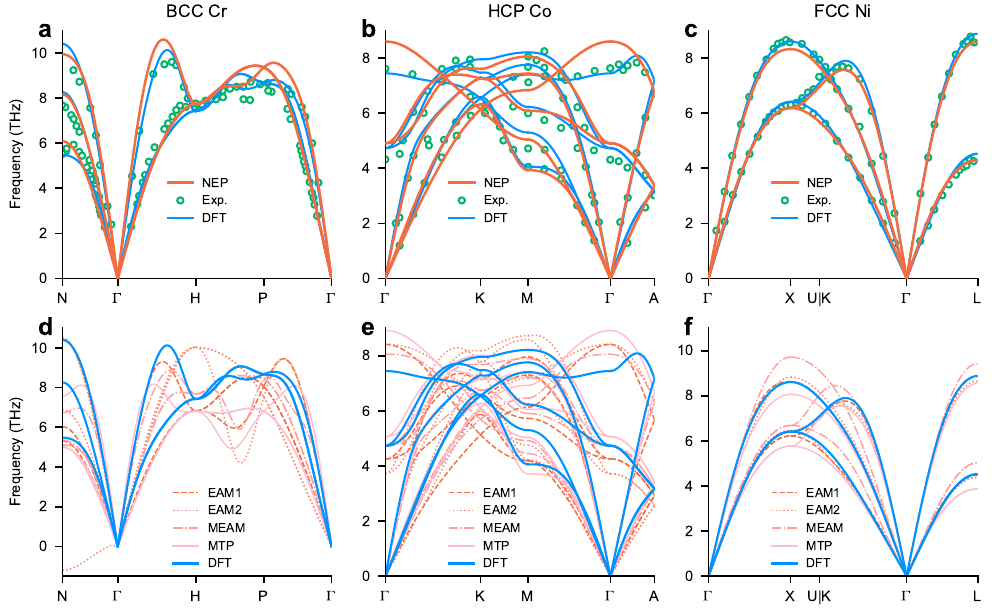}
    \caption{Phonon dispersions of (a) BCC Cr, (b) HCP Co, and (c) FCC Ni calculated using DFT, NEP, and experimental data. The corresponding comparisons with EAM1~\cite{li2019strengthening}, EAM2~\cite{farkas2020model}, MEAM~\cite{choi2018understanding}, and MTP~\cite{cao2025capturing} are shown in (d)–(f) for clarity. Experimental data for BCC Cr~\cite{shaw1971investigation}, HCP Co~\cite{wakaba1982lattice}, and FCC Ni~\cite{birgeneau1964normal}, obtained from room-temperature neutron diffraction measurements, are included for comparison.}
    \label{fig:phonon}
\end{figure}

Fig.~\ref{fig:phonon} compares the phonon dispersion relations of the elemental metals obtained from interatomic potentials and DFT calculations, together with available experimental data, to characterize lattice vibrational properties. The phonon spectra are computed using the finite-displacement method~\cite{phonopy} along high-symmetry directions in the Brillouin zone. The NEP results agree well with both DFT and experimental data for all three elemental metals, with only minor deviations observed for the optical phonon branches of Co. In contrast, other potentials show significant discrepancies for BCC Cr and HCP Co. In particular, EAM2 predicts imaginary phonon frequencies, indicating dynamical instability, as shown in Fig.~\ref{fig:phonon}d. Most potentials reproduce the phonon dispersions of Ni reasonably well; however, MEAM and MTP exhibit noticeable deviations in the high-frequency region (Fig.~\ref{fig:phonon}f). Overall, NEP provides the most accurate description of the phonon dispersions among the considered potentials.

We further compare several key crystal properties of the elemental metals and the ternary alloy, including the lattice parameter ($a$), elastic constants ($C_{ij}$), surface formation energy ($E_s$), and vacancy formation energy ($E_{vf}$), as summarized in Table~\ref{tab:basic}. Relative errors exceeding 25\% with respect to the DFT reference values are highlighted. All properties predicted by NEP fall within this threshold, demonstrating consistently high accuracy. In contrast, 8, 10, 10, and 7 properties predicted by EAM1, EAM2, MEAM, and MTP, respectively, exhibit errors larger than 25\%. All potentials predict equilibrium lattice constants that are comparable to the DFT and experimental values. For the surface energies of Ni and Co, all models give results consistent with DFT and correctly indicate that the \{111\} close-packed plane has the lowest surface energy in Ni, while \{10$\bar{1}$0\} corresponds to the highest surface energy in Co. In contrast, EAM2, MEAM, and MTP significantly underestimate the surface energy of Cr. Most potentials reasonably reproduce the vacancy formation energies of Ni and Co, while they generally underestimate this property in Cr. Notably, NEP also accurately reproduces the low-index grain boundary energies of Ni, predicting $\Sigma 3 [1\bar{1}0](111)$ as the lowest grain boundary energy, consistent with most experimental observations~\cite{randle2006mechanisms}. No grain boundary structures were included in the NEP training dataset, indicating strong transferability. In contrast, empirical potentials systematically overestimate the $\Sigma 3 [1\bar{1}0](112)$ grain boundary energy, while MTP even predicts a negative value for the $\Sigma 3 [1\bar{1}0](111)$ boundary. Additionally, all potentials yield reasonable intrinsic stacking fault energies ($\gamma_{\mathrm{isf}}$) for Ni except MTP, which predicts a negative $\gamma_{\mathrm{isf}}$, far from the DFT and experimental values. All empirical potentials also overestimate the unstable stacking fault energy ($\gamma_{\mathrm{usf}}$) of Ni. These results suggest that the empirical potentials and MTP are less reliable for describing defect-related properties in elemental systems, whereas NEP exhibits overall excellent performance for these key static properties.

\begin{table*}
\caption{Comparison of fundamental crystal properties for FCC Ni, BCC Cr, HCP Co, and equiatomic FCC CrCoNi alloy obtained from density functional theory (DFT), experiment (Exp.), and predicted by EAM1~\cite{li2019strengthening}, EAM2~\cite{farkas2020model}, MEAM~\cite{choi2018understanding}, MTP~\cite{cao2025capturing}, and NEP. These properties include the lattice parameter ($a$), elastic constants ($C_{ij}$), surface formation energy ($E_s$), vacancy formation energy ($E_{vf}$), low-index grain boundary energy, intrinsic stacking fault energy ($\gamma_{\mathrm{isf}}$), and unstable stacking fault energy ($\gamma_{\mathrm{usf}}$). Relative errors exceeding 25\% with respect to the DFT reference values are marked. DFT results from this work are used as the primary reference; when unavailable, DFT values from the literature are adopted.}
\label{tab:basic}
\centering
\begin{ruledtabular}
\begin{tabular}{ccccccccc}
         Property & DFT & DFT (other works)& Exp. & NEP & EAM1 & EAM2 & MEAM & MTP\\
        \midrule
        \multicolumn{9}{c}{FCC Ni} \\
        \midrule
         $a$ (\AA) &  3.52& 3.52~\cite{gong2024accurate} & 3.52~\cite{kanhe2016investigation} & 3.52 & 3.52 & 3.52 & 3.52 & 3.51\\
         $C_{11}$ (GPa) & 276 & 276~\cite{gong2024accurate} & 261~\cite{gong2024accurate} & 281 & 257 & 255 & 261 & 253 \\
         $C_{12}$ (GPa) & 158 & 156~\cite{gong2024accurate} & 151~\cite{gong2024accurate} & 162 & 151 & 151 & 151 & 163\\
         $C_{44}$ (GPa) & 129 & 131~\cite{gong2024accurate} & 132~\cite{gong2024accurate} & 124 & 122 & 134 & 132 & 110\\
         $E_{s}$\{100\} (J/$m^2$) &  2.27 & 2.24~\cite{gong2024accurate} & - & 2.22 & 2.20 & 1.88 & 1.94 & 1.76\\
         $E_{s}$\{110\} (J/$m^2$) &  2.30 & 2.34~\cite{gong2024accurate} & - & 2.28 & 2.39 & 2.05 & 2.04 & 1.86 \\
         $E_{s}$\{111\} (J/$m^2$) & 1.98 & 1.92~\cite{gong2024accurate} & - & 1.99 & 2.03 & 1.63 & 1.60 & 1.57\\
         $E_{vf}$ (eV) & - & 1.42~\cite{gong2024accurate}& 1.4--1.8~\cite{megchiche2006density} & 1.46 & 1.93$^{(+35.9\%)}$ & 1.60 & 1.51 & 1.38\\

         $\Sigma 3 [1\bar{1}0](111)$ (mJ/m$^2$) & - & 68~\cite{gong2024accurate} & - & 52 & 80 & 63 & 63 & -23$^{(-133.8\%)}$\\
         $\Sigma 3 [1\bar{1}0](112)$ (mJ/m$^2$) & - & 896~\cite{gong2024accurate} & - & 1029 & 1171$^{(+30.7\%)}$ & 1236$^{(+37.9\%)}$ & 1200$^{(+33.9\%)}$ & 872\\
         $\Sigma 5 [100](0\bar{2}1)$ (mJ/m$^2$) & - & 1289~\cite{gong2024accurate} & - & 1363 & 1429 & 1602 & 1449 & 1232\\
         $\Sigma 11 [1\bar{1}0](113)$ (mJ/m$^2$) & - & 454~\cite{gong2024accurate} & - & 447 & 535 & 541 & 516 & 379\\
         $\gamma_{isf}$ (mJ/m$^2$) & - & 136~\cite{gong2024accurate} & 125~\cite{carter1977the} & 105 & 160 & 129 & 126 & -46$^{(-133.8\%)}$\\
         $\gamma_{usf}$ (mJ/m$^2$) & - & 280~\cite{gong2024accurate} & - & 328 & 359$^{(+28.2\%)}$ & 403$^{(+43.9\%)}$ & 521$^{(86.1\%)}$ & 233\\

        \midrule
        \multicolumn{9}{c}{BCC Cr} \\
        \midrule
         $a$ (\AA) & 2.84 & 2.85~\cite{razumovskiy2011firstprinciples} & 2.88~\cite{gao2013phase} & 2.84 & 2.92 & 2.79 & 2.88 & 2.85\\
         $C_{11}$ & 487 & 484~\cite{razumovskiy2011firstprinciples}, 429~\cite{soderlind1994crystal} & 391~\cite{soderlind1994crystal} & 513 & 360$^{(-26.1\%)}$ & 171$^{(-64.9\%)}$ & 345$^{(-29.2\%)}$ & 552\\
         $C_{12}$ & 147 & 140~\cite{razumovskiy2011firstprinciples}, 119~\cite{soderlind1994crystal} & 90~\cite{soderlind1994crystal} & 164 & 172 & 170 & 114 & 381$^{(159.2\%)}$\\
         $C_{44}$ & 91 & 105~\cite{razumovskiy2011firstprinciples}, 91~\cite{soderlind1994crystal} & 103~\cite{soderlind1994crystal} & 95 & 76 & 137$^{(+50.5\%)}$ & 131$^{(+44.0\%)}$ & 66$^{(-27.5\%)}$\\
         $E_{s}$\{100\} (J/$m^2$) &  3.27 & - & - & 3.53 & 3.34 & 1.89$^{(-42.2\%)}$ & 2.37$^{(-27.5\%)}$ & 2.58\\
         $E_{s}$\{110\} (J/$m^2$) &  3.42 & - & - & 3.00 & 3.13 & 1.76$^{(-48.5\%)}$ & 2.43$^{(-28.9\%)}$ & 2.39$^{(-30.1\%)}$\\
         $E_{s}$\{111\} (J/$m^2$) &  3.49 & - & - & 3.59 & 3.43 & 2.11$^{(-39.5\%)}$ & 2.57$^{(-26.4\%)}$ & 2.76\\
         $E_{vf}$ (eV) & - & 2.86~\cite{korhonen1995vacancy} & - & 2.61 & 2.02$^{(-29.4\%)}$ & 0.51$^{(-82.2\%)}$ & 1.96$^{(-31.5\%)}$ & 2.10$^{(-26.6\%)}$\\
         
         \midrule
        \multicolumn{9}{c}{HCP Co} \\
        \midrule
         $a$ (\AA) & 2.49 & 2.52~\cite{kong2015phase}, 2.49~\cite{bideault2024polyvalent} & 2.51~\cite{dewaele2008compression} & 2.49 & 2.51 & 2.50 & 2.50 & 2.50\\
         $c$ (\AA) & 4.02 & 4.08~\cite{kong2015phase}, 4.02~\cite{bideault2024polyvalent} & 4.08~\cite{dewaele2008compression} & 4.02 & 4.09 & 4.20 & 4.08 & 4.00\\
         $C_{11}$ & 361 & 373~\cite{kong2015phase}, 386~\cite{bideault2024polyvalent} & 293~\cite{antonangeli2004elasticity} & 367 & 336 & 326 & 322 & 363\\
         $C_{12}$ & 153 & 166~\cite{kong2015phase}, 151~\cite{bideault2024polyvalent} & 143~\cite{antonangeli2004elasticity} & 171 & 191 & 182 & 140 & 151\\
         $C_{13}$ & 113 & 113~\cite{kong2015phase}, 114~\cite{bideault2024polyvalent} & 90~\cite{antonangeli2004elasticity} & 121 & 194$^{(+71.7\%)}$ & 110 & 123 & 94 \\
         $C_{33}$ & 426 & 386~\cite{kong2015phase}, 403~\cite{bideault2024polyvalent} & 339~\cite{antonangeli2004elasticity} & 385 & 354 & 396 & 339 & 319$^{(-25.1\%)}$\\
         $C_{44}$ & 100 & 93~\cite{kong2015phase}, 97~\cite{bideault2024polyvalent} & 78~\cite{antonangeli2004elasticity} & 76 & 58$^{(-42.0\%)}$ & 45$^{(-55.0\%)}$ & 69$^{(-31.0\%)}$ & 76\\
         $E_{s}$\{0001\} (J/$m^2$) &  2.08 & - & - & 2.28 & 2.18 & 1.60 & 1.93 & 1.70\\
         $E_{s}$\{10$\bar{1}$0\} (J/$m^2$) & 2.92 & - & - & 2.87 & 2.87 & 2.60 & 2.46 & 2.39\\
         $E_{s}$\{10$\bar{1}$1\} (J/$m^2$) &  2.08 & - & - & 2.28 & 2.18 & 1.60 & 1.93 & 1.70\\
         $E_{s}$\{11$\bar{2}$2\} (J/$m^2$) &  2.06 & - & - & 2.28 & 2.21 & 1.61 & 1.94 & 1.70\\
         $E_{vf}$ (eV) & 1.85 & - & 1.4~\cite{bideault2024polyvalent} & 1.67 & 1.52 & 1.47 & 1.47 & 1.64\\

         \midrule
        \multicolumn{9}{c}{FCC CrCoNi} \\
        \midrule
         $a$ (\AA) & - & 3.53~\cite{ge2018effect} & 3.56~\cite{laplanche2020processing} & 3.54 & 3.59 & 3.53 & 3.55 & 3.53 \\
         $C_{11}$ & - & 267~\cite{ge2018effect} & 249~\cite{laplanche2020processing} & 295 & 235 & 215 & 247 & 303  \\
         $C_{12}$ & - & 182~\cite{ge2018effect} & 159~\cite{laplanche2020processing} & 185 & 176 & 146 & 165 & 191  \\
         $C_{44}$ & - & 178~\cite{ge2018effect} & 138~\cite{laplanche2020processing} & 142 & 90$^{(-49.4\%)}$ & 130$^{(-27.0\%)}$ & 83$^{(-53.4\%)}$ & 151 \\
         
\end{tabular}
\end{ruledtabular}
\end{table*}

\subsection{Short range order}

\begin{figure}[H]
    \centering
    \includegraphics[width=0.8\linewidth]{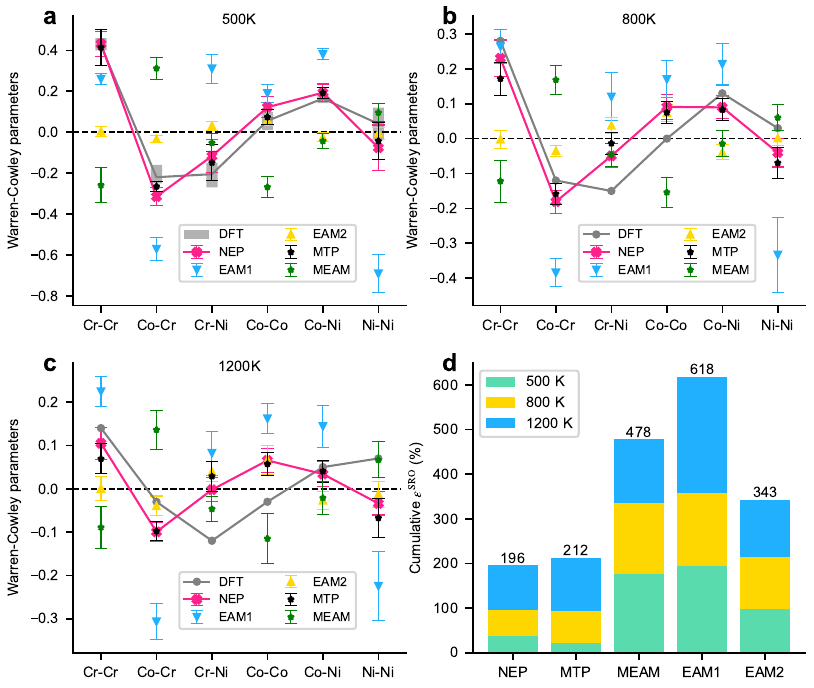}
    \caption{Comparison of Warren–Cowley short-range order parameters at (a) 500 K, (b) 800 K, and (c) 1200 K as predicted by NEP, EAM1~\cite{li2019strengthening}, EAM2~\cite{farkas2020model}, MEAM~\cite{choi2018understanding}, and MTP~\cite{cao2025capturing}. Error bars represent the standard deviation obtained from 20 independent simulations with different initial configurations. At 500 K, the shaded gray region denotes the range of DFT MC results reported in three literature sources~\cite{cao2025capturing, ding2018tunable, tamm2015atomic}. For 800 and 1200 K, where only single DFT reference values are available~\cite{tamm2015atomic}, the results are shown as discrete points. (d) Cumulative relative error ($\varepsilon^{\mathrm{SRO}}$) with respect to DFT.}
    \label{fig:mcmd}
\end{figure}

It has been established that chemical short-range order affects various chemistry-microstructure relationships that influence the mechanical properties of alloys. The SRO in crystalline phases can be quantified using the Warren-Cowley (WC) parameters $\alpha_{ij} = 1 - p_{ij}/c_{j}$, where $p_{ij}$ denotes the probability of finding an atom of species $j$ in the first-neighbor shell of an atom of type $i$, and $c_j$ is the atomic concentration of species $j$. A value of $\alpha_{ij}=0$ corresponds to a completely random solution. Positive values of $\alpha_{ij}$ indicate a tendency to reduce the number of $i$-$j$ atomic pairs, whereas negative values indicate a preference for forming $i$-$j$ pairs. Here, we compute the SRO of equiatomic CrCoNi MEA at various temperatures using a hybrid MC/MD strategy (see details in the Methods section) with different interatomic potentials, and compare the results with DFT-based MC simulations. As shown in Fig.~\ref{fig:mcmd}, both NEP and MTP reproduce the WC parameters well at 500~K. In contrast, MEAM only captures the tendencies of Cr--Ni and Ni--Ni pairs, while EAM1 reproduces only the Co--Co pair tendency and fails for the other pairs. Notably, EAM2 predicts WC parameters close to zero for all pairs, suggesting the absence of SRO, which is physically unreasonable. With increasing temperature, the absolute values of the WC parameters generally decrease, indicating weaker SRO effects. This trend is consistent with previous computational studies~\cite{cao2025capturing, ghosh2022short}. Although some discrepancies between NEP and DFT results appear at 1200~K, the overall variation trends remain similar. To quantitatively evaluate the accuracy of the SRO predictions for different potentials, the relative error with respect to the DFT reference is defined as $\varepsilon^{\mathrm{SRO}}=(\sum_{i=1}^{3}\sum_{j=i}^{3}\left|\alpha_{ij}^{\mathrm{MD}}-\alpha_{ij}^{\mathrm{DFT}}\right|)/
    (\sum_{i=1}^{3}\sum_{j=i}^{3}\left|\alpha_{ij}^{\mathrm{DFT}}\right|)$. The cumulative $\varepsilon^{\mathrm{SRO}}$ for the three temperatures is shown in Fig.~\ref{fig:mcmd}d. NEP exhibits the lowest error, while MTP shows comparable performance and significantly smaller errors than the empirical potentials. These results demonstrate that NEP can reliably reproduce SRO behavior over a wide temperature range.

\subsection{Stacking fault energy and dislocation dissociation}

\begin{figure}[H]
    \centering
    \includegraphics[width=0.85\linewidth]{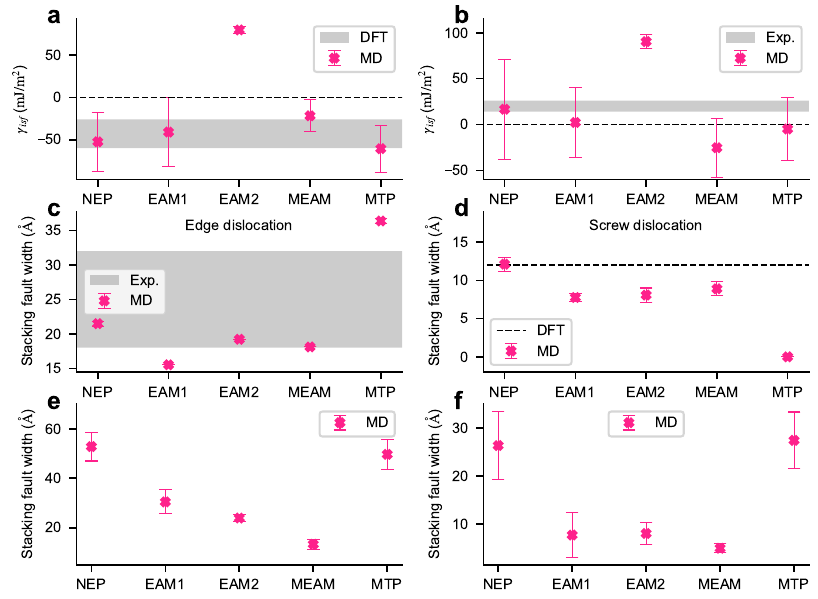}
    \caption{Intrinsic stacking fault energies $\gamma_{\mathrm{isf}}$ for (a) random solid solutions and (b) ordered solid solutions, predicted by NEP, EAM1~\cite{li2019strengthening}, EAM2~\cite{farkas2020model}, MEAM~\cite{choi2018understanding}, and MTP~\cite{cao2025capturing}. DFT results for random solid solutions in (a) are taken from Ref.~\cite{zhu2023effects}, while experimental data for ordered solid solutions in (b) are from Refs.~\cite{laplanche2017reasons, liu2018stacking}. Stacking fault widths associated with edge and screw dislocations in (c–d) Ni and (e–f) CrCoNi. The experimental data in (c) are taken from Ref.~\cite{carter1977the}, while the DFT results in (d) are from Ref.~\cite{tan2019dislocation}.}
    \label{fig:sfe_dislo}
\end{figure}

Based on the validation of the model’s ability to capture short-range order in the previous section, we further investigate its effects on the stacking fault energy of the CrCoNi alloy. Fig.~\ref{fig:sfe_dislo} presents the SFE of both random and ordered CrCoNi solid solutions equilibrated using hybrid MC/MD simulations using various potentials. For each case, the results are averaged over 150 independent configurations to ensure the robustness of the predicted SFE; additional details are provided in the Methods section. For the random solid solution, the NEP-predicted SFE is -52 mJ/m$^2$, which falls well within the DFT range of -26 to -60 mJ/m$^2$, as summarized in Ref.~\cite{zhu2023effects}. This wide variation suggests that SFE calculations require sampling a sufficiently large number of configurations to achieve convergence, whereas only a limited number of structures have been considered in some previous DFT studies~\cite{zhang2017dislocation, zhang2017origin}. Note that ~\citeauthor{ding2018tunable}~\cite{ding2018tunable} calculated the SFE using 108 configurations and reported an average DFT value of -43 mJ/m$^2$, which is close to our prediction. Among the empirical potentials, MTP slightly underestimates the SFE (-61 mJ/m$^2$), whereas MEAM slightly overestimates it (-21 mJ/m$^2$). EAM1 yields a reasonable average SFE (-41 mJ/m$^2$), in contrast, EAM2 predicts a positive SFE (80 mJ/m$^2$), which is physically inconsistent with both DFT calculations and previous simulations. Despite these discrepancies, both MD and DFT calculations consistently predict a negative SFE for the random alloy, which contradicts experimental measurements discussed earlier. 

To resolve this discrepancy, we further examine the influence of chemical SRO. As shown in Fig.~\ref{fig:sfe_dislo}b, the SFE of the ordered alloy predicted by NEP increases significantly from -52 to 17 mJ/m$^2$, placing it within the experimental range of 18$\pm$4 mJ/m$^2$~\cite{liu2018stacking} and 22$\pm$4 mJ/m$^2$~\cite{laplanche2017reasons}. This result is also highly consistent with the DFT-MC study of ~\citeauthor{ding2018tunable}~\cite{ding2018tunable}, in which the SFE increases from -43 to 30 mJ/m$^2$, averaged over 18 initial structures. These findings support the conclusion that chemical SRO can significantly increase the SFE and suggest that experimentally studied CrCoNi alloys likely possess a substantial degree of local chemical ordering rather than existing as purely random solid solutions. EAM1 and MTP exhibit a similar trend of increasing SFE with the introduction of SRO, although they still underestimate the SFE of the ordered structures. In contrast, EAM2 and MEAM show only minor changes in SFE, indicating that these potentials are unable to adequately capture the effects of SRO in the CrCoNi alloy. 

It is well established that stacking faults are generated during the dissociation of lattice dislocations into partial dislocations in order to reduce the elastic energy~\cite{shih2021stacking}. The width of the stacking fault region, defined as the distance between the two partial dislocations, is therefore determined by the balance between the repulsive elastic interaction of the partial dislocations and the energy required to form the stacking fault between them, i.e., SFE. Fig.~\ref{fig:sfe_dislo}c-f show the stacking fault width (SFW) resulting from the dissociation of edge and screw dislocations in Ni and CrCoNi at~0~K. For the edge dislocation in Ni, the SFW predicted by NEP is 21.5 \AA, which agrees well with the experimental measurement of $26\pm8$ \AA~\cite{carter1977the} (see Fig.~\ref{fig:sfe_dislo}c). EAM2 and MEAM yield SFW values of 19.2 and 18.2 \AA, respectively, both of which also fall within the experimental range, whereas EAM1 underestimates the SFW, giving a value of 15.5 \AA. These results exhibit an inverse correlation with the SFE values (see Table~\ref{tab:basic}), as expected from classical elastic theory~\cite{bacon1980anisotropic}. Notably, MTP predicts a significantly larger SFW of 36.4 \AA, which can be attributed to its unphysical negative SFE. For the screw dislocation in Ni, the SFW predicted by NEP is 12.1 \AA. Although experimental measurements for this quantity are scarce, this value is in good agreement with previous DFT calculations, which report an SFW of approximately 12 \AA~\cite{tan2019dislocation}. In contrast, all empirical potentials underestimate this value, while MTP predicts no dissociation (SFW = 0), as shown in Fig.~\ref{fig:sfe_dislo}d.

For the CrCoNi alloy, although the SFE is negative, the SFW remains finite at 0 K, consistent with previous theoretical studies~\cite{shih2021stacking}. The SFWs for both edge and screw dislocations are larger than those in Ni due to the lower SFE. NEP and MTP produce similar average results and successfully capture fluctuations arising from variations in the local chemical environment; the reported values are averaged over ten independent configurations. In contrast, EAM2 and MEAM show nearly identical results across different structures, indicating a lack of sensitivity to chemical disorder. While EAM1 partially captures this variability, it systematically underestimates the SFW compared to NEP and MTP. The effects of temperature and SRO on dislocation dissociation require substantially greater computational effort and will be addressed in future work. In comparing zero-temperature calculations with room-temperature experiments, finite-temperature stacking fault energies must include entropy contributions from atomic vibrations, magnetism, and electronic excitations~\cite{niu2018magnetically}. Niu \textit{et al.}~\cite{niu2018magnetically} showed that vibrational contributions change the HCP-FCC free-energy difference by less than 15\% between 0 and 300 K in random CoCrNi alloys. We therefore expect a similarly minor effect on SFE over this temperature range, implying that experimental values extrapolated to 0 K would be only slightly reduced. Accordingly, the SFE-dependent SFW should also exhibit only minor changes over this temperature range.

\subsection{Composition-dependent properties of CrCoNi alloys}

\begin{figure}[H]
    \centering
    \includegraphics[width=0.85\linewidth]{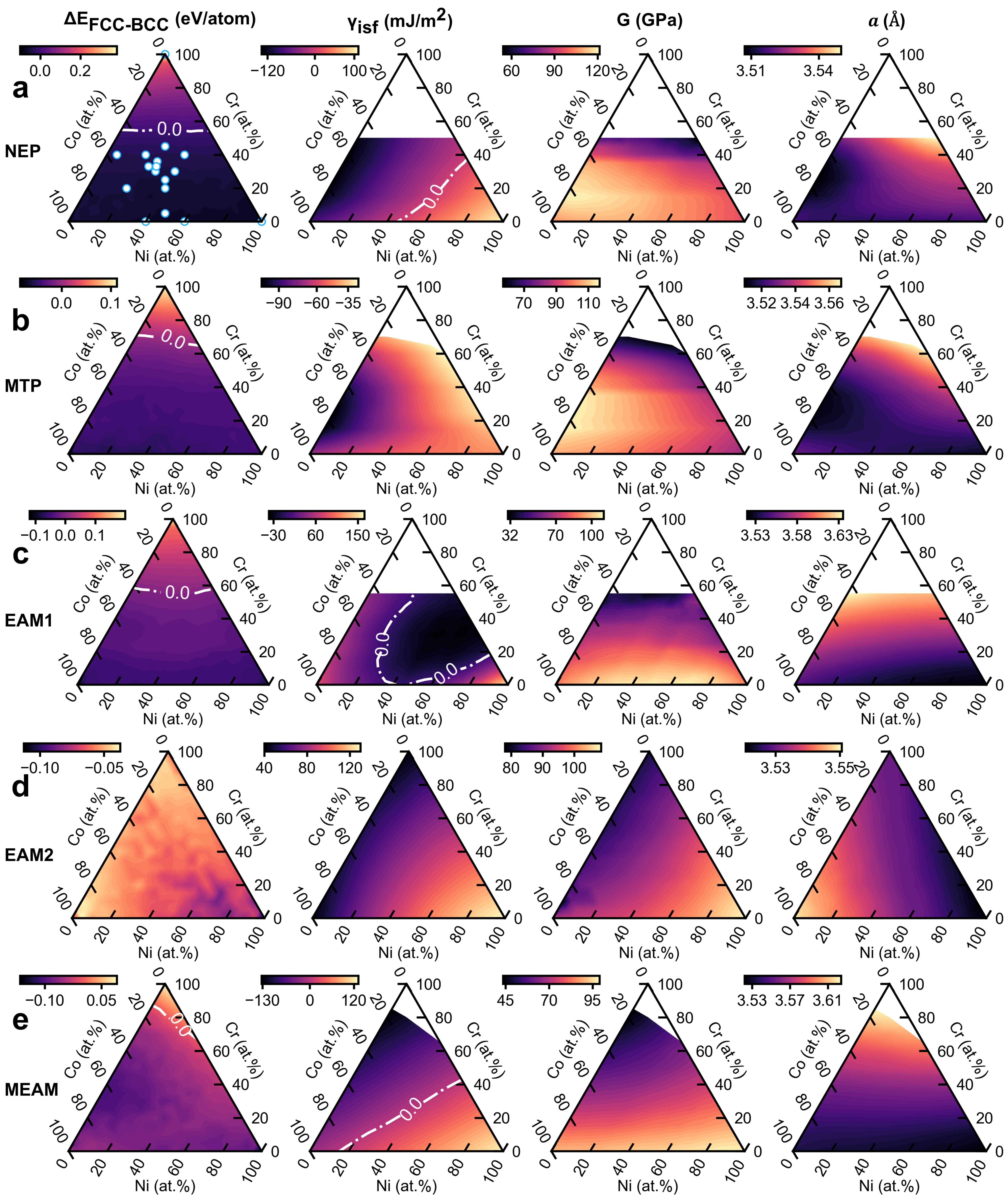}
    \caption{Ternary plots of $\Delta E_{\mathrm{FCC-BCC}}$, intrinsic stacking fault energy $\gamma_{\mathrm{isf}}$, shear module $G$ and lattice constant $a$ predicted by (a) NEP, (b) MTP~\cite{cao2025capturing}, (c) EAM1~\cite{li2019strengthening}, (d) EAM2~\cite{farkas2020model} and (e) MEAM~\cite{choi2018understanding} across the CrCoNi compositional space. White symbols in (a) indicate experimentally observed stable FCC compositions, including Cr$_{30}$Co$_{40}$Ni$_{30}$~\cite{deng2021enhancement}, Cr$_{40}$Co$_{55}$Ni$_{5}$~\cite{coury2019design}, Cr$_{45}$Co$_{27.5}$Ni$_{27.5}$~\cite{coury2018highthrou}, Cr$_{25}$Co$_{37.5}$Ni$_{37.5}$~\cite{coury2018highthrou}, Cr$_{30}$Co$_{30}$Ni$_{40}$~\cite{yang2022theoretical}, Cr$_{36}$Co$_{36}$Ni$_{28}$~\cite{yang2022theoretical}, Cr$_{40}$Co$_{20}$Ni$_{40}$~\cite{shuhei2020effectof}, Cr$_{20}$Co$_{60}$Ni$_{20}$~\cite{shuhei2020effectof}, Cr$_{33}$Co$_{38}$Ni$_{29}$~\cite{huang2022advanced}, Cr$_{33}$Co$_{42}$Ni$_{25}$~\cite{huang2022advanced}, Cr$_{40}$Co$_{40}$Ni$_{20}$~\cite{gustavo2021hallpetch}, Co$_{40}$Ni$_{60}$~\cite{madhavan2014texture}, Co$_{60}$Ni$_{40}$~\cite{yoshida2019deformation}, Cr$_{5}$Co$_{47.5}$Ni$_{47.5}$~\cite{yoshida2019deformation}, Cr$_{20}$Co$_{40}$Ni$_{40}$~\cite{yoshida2019deformation}, as well as pure FCC Ni and pure BCC Co.}
    \label{fig:tenary}
\end{figure}

We further examine the composition-dependent FCC–BCC phase stability, SFE, shear modulus $G$, and lattice constant $a$ across the CrCoNi ternary space, as shown in Fig.~\ref{fig:tenary}. The NEP model predicts that the FCC phase is more stable than the BCC phase when the Cr content is below ~55 at.\%, where $\Delta E_{\mathrm{FCC-BCC}} < 0$. This agrees well with experimental observations, in which stable FCC phases are predominantly found in this compositional regime (Fig.~\ref{fig:tenary}a). EAM1 yields similar results, whereas MTP and MEAM overestimate the Cr content at the FCC–BCC phase boundary. In contrast, EAM2 produces an unphysical prediction, with the FCC phase more stable than BCC across the entire composition space. Only compositions corresponding to stable FCC phases are considered for evaluating SFE, $G$, and $a$. The NEP model predicts a strong composition dependence of the SFE, which increases from negative to positive values with increasing Ni content and decreasing Cr content, consistent with first-principles calculations. For example, Yan \textit{et al.}~\cite{yan2023design} reported SFE values ranging from –46.4 to 88.5 mJ/m$^2$ as the Cr content decreases from 67 to 0 at.\% in Co$_{33}$Cr$_x$Ni$_{67-x}$ alloys, and predicted an FCC–BCC transition at $x < 57.3$, in good agreement with NEP. Although MTP reproduces a similar trend, it underestimates the SFE and fails to yield positive values, likely due to its inaccurate description of pure Ni, indicating limited transferability to non-equimolar compositions. EAM1 and EAM2 show significant deviations, with the former mispredicting SFE at low Ni content and the latter yielding uniformly positive SFE. MEAM captures the qualitative trend but predicts a significantly shifted transition boundary.

For elastic properties, NEP predicts that $G$ increases from ~60 to 120 GPa with increasing Co content and decreasing Ni/Cr content, consistent with the higher intrinsic stiffness of Co (Table~\ref{tab:basic}). This trend agrees well with first-principles results~\cite{yan2023design}, which show an increase in $G$ from 86.5 to 123.4 GPa as Co content increases from 0 to 67 at.\% in Co$_x$Cr$_{33}$Ni$_{67-x}$. MTP reproduces a similar trend, whereas other empirical potentials fail to capture it accurately. The lattice constant $a$ is also well reproduced by NEP, decreasing from 3.55 to 3.51 \AA ~with increasing Co content, in close agreement with first-principles calculations~\cite{yan2023design}. Overall, these results demonstrate that the NEP model provides an accurate and consistent description of composition-dependent properties across both equimolar and non-equimolar CrCoNi alloys, outperforming existing potentials.

\subsection{Melting properties}

\begin{figure}[H]
    \centering
    \includegraphics[width=0.85\linewidth]{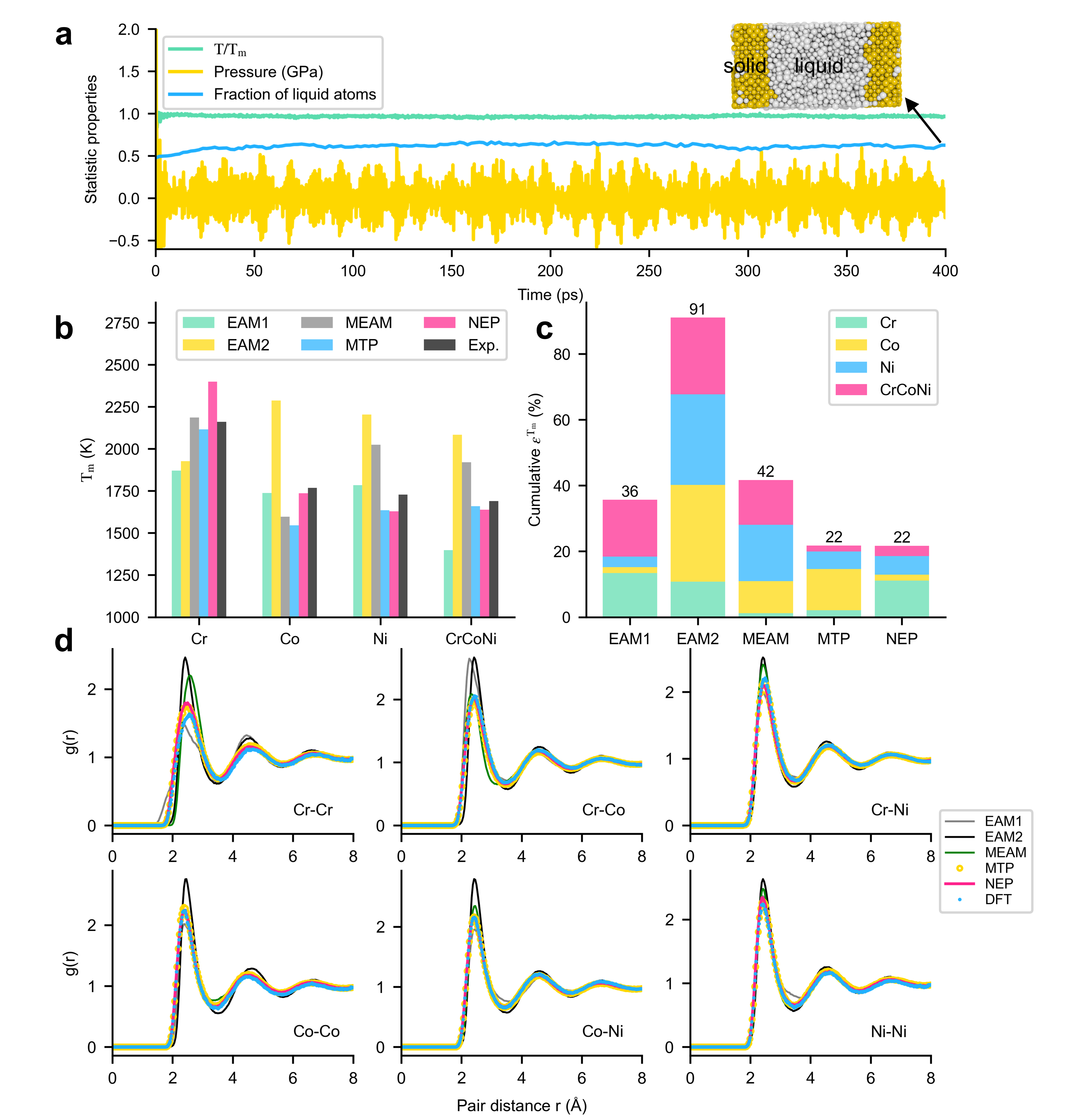}
    \caption{(a) Evolution of normalized temperature, pressure, and liquid fraction during the two-phase coexistence process of the CrCoNi alloy computed using NEP. The inset shows the atomic configuration at 400 ps, colored by PTM analysis~\cite{larsen2016robust}, where yellow represents FCC atoms and grey indicates liquid atoms.
(b) Melting temperature ($\mathrm{T_m}$) predicted by EAM1~\cite{li2019strengthening}, EAM2~\cite{farkas2020model}, MEAM~\cite{choi2018understanding}, MTP~\cite{cao2025capturing}, and NEP.
(c) Cumulative relative error $\varepsilon^{\mathrm{T_m}}$, compared with experimental values for Cr~\cite{ioroi2022melting}, Co~\cite{bideault2024polyvalent}, Ni~\cite{gong2024accurate}, and CrCoNi~\cite{wu2014temperature}.
(d) Partial radial distribution functions g(r) of the liquid phase for equiatomic CrCoNi at 2684 K, calculated using potentials and DFT~\cite{cao2025capturing}.}
    \label{fig:liquid}
\end{figure}

We further evaluate the performance of the NEP model under high-temperature conditions by examining its ability to predict the melting points of elemental metals and the equiatomic CrCoNi, as well as its capability to describe the partial radial distribution functions g(r) of the alloy in the fully liquid state.
The melting temperature ($\mathrm{T_m}$) is evaluated using solid–liquid phase coexistence molecular dynamics simulations, in which the temperature, pressure, and fraction of liquid atoms remain stable for 400 ps, as shown in Fig.~\ref{fig:liquid}a. The inset showing the microstructure of CrCoNi in the final simulation frame further illustrates the clear interface between the solid and liquid phases. The melting temperature is obtained by averaging the temperature over the last 200 ps of the simulation.

The predicted $\mathrm{T_m}$ values from different potentials are compared with experimental data in Fig.~\ref{fig:liquid}b. To quantitatively assess the accuracy, the relative error with respect to experimental measurements is defined as $\varepsilon^{\mathrm{T_m}} = (\mathrm{T_m^{MD}} - \mathrm{T_m^{Exp})/\mathrm{T_m^{Exp}}}$.
The NEP model slightly overestimates the melting temperature of Cr, whereas MTP underestimates that of Co. Nevertheless, both potentials show high accuracy for Ni and the CrCoNi alloy, and their cumulative $\varepsilon^{\mathrm{T_m}}$ values are comparable, as shown in Fig.~\ref{fig:liquid}c. In particular, the melting temperature of CrCoNi predicted by NEP is 1638 K, which is close to the experimental value of 1690 K~\cite{wu2014temperature}. In contrast, the empirical potentials exhibit significantly larger deviations in predicting the melting temperature of CrCoNi, yielding values of 1398 K (EAM1), 2084 K (EAM2), and 1920 K (MEAM).

The partial radial distribution functions of liquid CrCoNi at 2684 K are compared in Fig.~\ref{fig:liquid}d for all elemental pairs. The results show that the predictions from NEP and MTP almost completely overlap and both agree well with the DFT-based \textit{ab initio} molecular dynamics results~\cite{cao2025capturing}. The largest deviation occurs for the Cr–Cr pair, where the position of the first peak is slightly shifted to a shorter distance by approximately 0.05 \AA. Notably, the first peak of the Cr–Cr pair is significantly lower than those of the other elemental pairs, indicating that this chemical interaction is energetically less favorable. This observation suggests that the liquid phase is also influenced by the considerable chemical complexity present in the crystalline phase. In contrast, the other empirical potentials exhibit varying degrees of error and fail to reproduce the DFT radial distribution curves.

\subsection{Uniaxial tensile deformation}

\begin{figure}[H]
    \centering
    \includegraphics[width=1\linewidth]{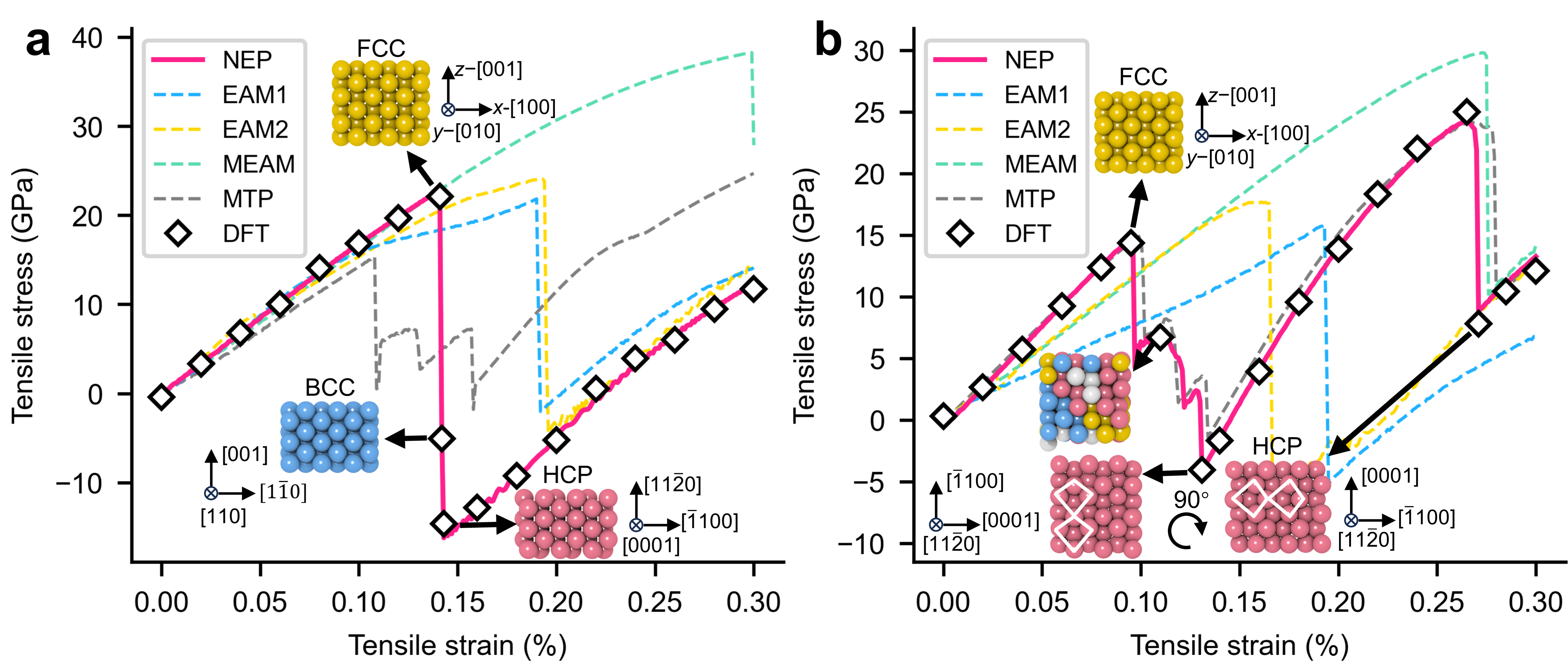}
    \caption{Stress–strain curves under uniaxial tensile deformation for (a) Ni and (b) CrCoNi predicted using EAM1~\cite{li2019strengthening}, EAM2~\cite{farkas2020model}, MEAM~\cite{choi2018understanding}, MTP~\cite{cao2025capturing}, NEP, and DFT. The inset structures are colored according to the local crystal structure identified by PTM analysis~\cite{larsen2016robust}, where FCC, BCC, HCP, and other structures are shown in yellow, blue, red, and grey, respectively.}
    \label{fig:deformation}
\end{figure}

As a stringent test of the NEP model, we probe the allotropic phase transformations of Ni and equiatomic CrCoNi under uniaxial tension along [100] at a constant strain rate of $10^9$ s$^{-1}$, using a $3 \times 3 \times 3$ conventional supercell (108 atoms). For pure Ni, the stress–strain response shows a monotonic increase in stress with increasing strain, followed by a sudden drop at a strain of approximately 0.14, as shown in Fig.~\ref{fig:deformation}a. The inset atomic configurations illustrate the corresponding atomic structures at different strain levels, revealing the evolution from strained FCC to BCC and finally to HCP structures. These snapshots indicate that the abrupt stress drop corresponds to an FCC–BCC–HCP phase transformation. The stress changes from tensile (positive) to compressive (negative) due to the volume change associated with the phase transformation. A similar phenomenon has also been reported in tensile-induced FCC–BCC transformations in HEA in a previous MD study~\cite{li2018transformation}. Subsequently, the HCP phase remains stable, and the stress increases again as the strain continues to 0.3.

A previous machine-learning potential developed specifically for Ni~\cite{gong2024accurate} also predicted an FCC-to-HCP transformation, following the same crystallographic orientation relationship, namely $\{010\}_{\mathrm{FCC}}\parallel\{0001\}_{\mathrm{HCP}}$ and $\langle 001 \rangle_{\mathrm{FCC}}\parallel\langle 11\bar{2}0 \rangle_{\mathrm{HCP}}$. The main difference is that our model predicts a transient BCC phase during the transformation. This intermediate phase exists only for a very short time and is likely related to the high strain rate and finite-size effects in the simulation. Similar FCC-BCC-HCP transformation pathways have been reported in previous MD studies of Cu nanowires~\cite{xie2015anew}, where the transformation was also induced by high strain rates. In that work, the orientation relationship was reported as $\{010\}_{\mathrm{FCC}} \parallel \{110\}_{\mathrm{BCC}} \parallel \{0001\}_{\mathrm{HCP}}$. Moreover, strain-induced FCC-to-HCP transformations in nanocrystalline Ni have also been observed experimentally~\cite{guo2021plastic, luo2019plastic}. The stress–strain behavior predicted by NEP is in excellent agreement with our DFT results, and the average atomic energy at the transition state decreases from -5.38 eV/atom (FCC) to -5.40 eV/atom (BCC) and further to -5.42 eV/atom (HCP), indicating that the HCP structure becomes energetically more stable under the applied strain. In contrast, the other potentials predict different mechanical responses and phase transformation pathways compared with the NEP and DFT results.

Fig.~\ref{fig:deformation}b shows the stress–strain response for the CrCoNi MEA. In this case, the transition state between the FCC and HCP phases exhibits a mixed-phase structure due to the complex chemical environment and lattice distortion. The FCC-to-HCP transformation follows the same crystallographic orientation relationship observed in Ni. Interestingly, a second stress drop occurs within the HCP phase at a strain of 0.265, followed by a recovery in stress as the strain increases to 0.3. The inset structures in Fig.~\ref{fig:deformation}b indicate that this behavior is associated with a rotation of the crystalline orientation during deformation. Notably, an \textit{in situ} neutron diffraction study reported a bulk FCC–HCP phase transformation in CrCoNi at 15 K under tensile loading, beginning at a strain of approximately 0.14~\cite{he2021stacking}. The stress predicted by NEP is very close to the DFT results, further demonstrating the high accuracy of the model. MTP also shows a similar tensile response to NEP, whereas the other empirical potentials fail to reproduce the observed mechanical deformation behavior.

\subsection{Computational efficiency}

\begin{figure}[H]
    \centering
    \includegraphics[width=0.85\linewidth]{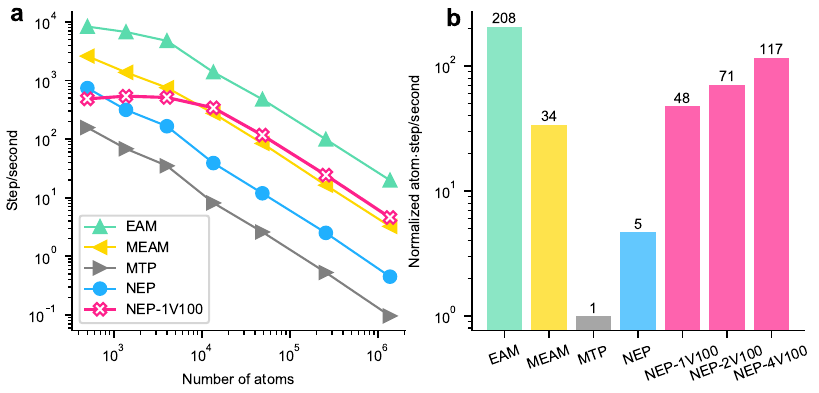}
    \caption{Benchmark of computational efficiency for different interatomic potentials, including EAM~\cite{li2019strengthening}, MEAM~\cite{choi2018understanding}, MTP~\cite{cao2025capturing}, and NEP. (a) Simulation speed (steps per second) as a function of system size. (b) Computational throughput (atoms $\times$ steps per second) for the different potentials, normalized to the performance of MTP for a system containing 1,372,000 atoms. All CPU-based tests were performed on a 64-core AMD Rome 7H12 processor using LAMMPS (22 Jul 2025 Update 2)~\cite{lammps}. For NEP, GPU performance on one to four Tesla V100-SXM2-16GB devices using GPUMD-4.9~\cite{gpumd} is also presented.
}
    \label{fig:speed}
\end{figure}

In the above sections, we have examined the accuracy of the developed NEP model in predicting fundamental crystal properties, short-range order, stacking fault energies, high-temperature properties, and deformation behavior. However, despite the high accuracy of MLPs, their computational efficiency is often less discussed~\cite{cao2025capturing, gong2024accurate}. In general, MLPs are significantly faster than DFT calculations, as the computational cost of MLPs scales linearly with system size, whereas DFT typically scales cubically. Nevertheless, MLPs may still be considerably slower than classical empirical potentials, which can limit their applicability in large-scale and long-time molecular dynamics simulations. Here, we compare the computational efficiency of two MLPs, NEP and MTP, with classical empirical potentials, including EAM and MEAM. As shown in Fig.~\ref{fig:speed}a, all potentials exhibit approximately linear scaling with increasing system size, accompanied by a decrease in the number of simulation steps per second. However, the fastest potential, EAM, is nearly two orders of magnitude faster than the slowest potential, MTP. For the GPU-accelerated NEP model, the simulation speed remains nearly constant as the system size increases up to approximately 10,000 atoms, after which it decreases linearly. This behavior arises because the GPU becomes fully utilized beyond this system size. 

The computational throughput (atoms $\times$ steps per second) for a system containing 1,372,000 atoms is shown in Fig.~\ref{fig:speed}b. The performance is normalized with respect to MTP. NEP running on a CPU is approximately five times faster than MTP, while the use of a single Tesla V100 GPU provides a 48-fold speedup; notably, the GPU has a comparable cost to the CPU platform. A recently developed GPU-accelerated MTP implementation~\cite{meng2025akokkos} is not included in this comparison due to its substantially higher memory requirements. For the system considered here, it requires on the order of hundreds of gigabytes of GPU memory, whereas NEP can be executed with only 16 GB, highlighting its superior memory efficiency. Using four V100 GPUs in parallel further increases the speed to approximately 117 times that of MTP, with a parallel efficiency of 61\%. This level of acceleration enables large-scale molecular dynamics simulations with machine-learning potentials.

\section{Conclusion}

In conclusion, we developed a high-fidelity machine-learning potential based on the NEP framework for the CrCoNi system. The model achieves near–DFT-level accuracy in predicting a wide range of properties for both elemental metals and equiatomic or non-equiatomic CrCoNi medium-entropy alloys, including lattice parameters, elastic constants, equations of state, phonon spectra, defect energetics, dislocation dissociation, and melting temperatures. For the CrCoNi alloy, the NEP model accurately captures chemical short-range order across a range of temperatures and quantitatively reveals its influence on stacking fault energy, bringing the predicted values into agreement with experimental measurements. In addition, the model successfully describes strain-induced FCC–HCP phase transformations and their crystallographic orientation relationships during deformation. Comprehensive benchmarking against widely used empirical potentials and existing machine-learning potentials demonstrates that the NEP model provides superior overall accuracy while maintaining high computational efficiency. This capability enables reliable large-scale and long-time MD/MC simulations for the investigation and design of advanced CrCoNi MEAs. Furthermore, the dataset generated in this work provides a valuable foundation for the future development of MLPs for CrCoNi-based HEAs.

\section{Methods}

\subsubsection{Density functional theory calculations}

All DFT calculations were performed using the Perdew-Burke-Ernzerhof (PBE)~\cite{pberef} exchange-correlation functional and projector-augmented wave (PAW) pseudopotentials~\cite{pawref}, as implemented in the Vienna \textit{ab initio} Simulation Package (VASP, version 6.4.2)~\cite{vasp1, vasp2, vasp3, vasp4, vasp5}. The PAW datasets (potpaw.PBE.64) treated the valence electrons of Cr, Co, and Ni as 3p$^{6}$3d$^{5}$4s$^{1}$, 3d$^{8}$4s$^{1}$, and 3d$^{8}$4s$^{2}$, respectively. All calculations were carried out within a collinear spin-polarized framework, with an initial magnetic moment of $3~\mu_{\mathrm{B}}$ assigned to each element, where $\mu_{\mathrm{B}}$ denotes the Bohr magneton. Convergence tests were conducted separately for the plane-wave energy cutoff and the Monkhorst-Pack \textit{k}-point mesh density. These parameters were iteratively optimized until the total energy converged to within $1~\mathrm{meV/atom}$. The final converged settings were an energy cutoff of $450~\mathrm{eV}$ for all elemental systems, and a minimum \textit{k}-point spacing of $0.25~\mathrm{\AA^{-1}}$. Because the systems studied are metallic, a smearing scheme was employed to improve the convergence of the electronic self-consistency cycle. Specifically, first-order Methfessel-Paxton smearing with a width of $0.1~\mathrm{eV}$ was used. The electronic self-consistency and ionic relaxation convergence criteria were set to $10^{-5}~\mathrm{eV}$ and $0.02~\mathrm{eV/\AA}$, respectively.

\subsubsection{MLP development}

The development of the NEP model consists of two main stages: (i) preparation of the initial dataset and (ii) iterative exploration \textit{via} active learning. The initial dataset was constructed by combining two existing MTP datasets. The first dataset was developed by Cao \textit{et al.}~\cite{cao2025capturing} and includes the TS-0, TS-3, and TS-f subsets. It should be noted that the so-called “MTP dataset” referenced in Fig.~\ref{fig:dataset} comprises the complete set of TS-0 to TS-5 and TS-f; detailed descriptions of these subsets can be found in the Methods section of Ref.~\cite{cao2025capturing}. The second dataset was originally developed for the CrCoFeNi system, from which the CrCoNi subset was extracted~\cite{nitol2025achieving}.

Since these datasets lack sufficient coverage of non-equiatomic CrCoNi compositions, we further constructed additional structures by systematically sampling the composition space. Specifically, the concentration of each element was varied from 0 to 100 at.\% with an interval of 4 at.\%, and for each composition both FCC and BCC structures were generated. To enhance structural diversity, random atomic displacements of $\pm$0.1 \AA\ were applied. The resulting dataset contains approximately 8000 structures. To reduce redundancy and ensure representative sampling, we employed the periodic SOAP descriptor~\cite{alber2013onrepre} as implemented in the Dscribe package~\cite{dscribe2} and performed principal component analysis. The SOAP parameters were set as follows: a cutoff radius of 6.0 \AA, $n_{\mathrm{max}} = 8$, and $l_{\mathrm{max}} = 6$. Farthest point sampling was then applied in the PCA-reduced space to select 2000 representative structures, which were subsequently labeled using high-accuracy DFT calculations and used as the initial training dataset.

We then employed an iterative active learning strategy~\cite{wu2025revealing} to further enrich the dataset. Starting from the initial dataset, a preliminary NEP model was trained and subsequently used to perform MD simulations to sample new configurations for retraining. During this process, we progressively incorporated diverse configurations, including elemental systems sampled over a temperature range of 1–3000 K, point-defect structures at 300–1000 K, stacking faults, as well as structures subjected to compressive and tensile deformations. All thermal simulations were conducted under zero-pressure conditions. The systems considered include FCC Ni, BCC Cr, HCP Co, and binary and ternary FCC CrCoNi alloys. At each iteration, configurations that were significantly outside the existing training domain were discarded, and structures associated with non-converged DFT calculations were excluded. The latter is commonly encountered in spin-polarized magnetic calculations and is computationally demanding. After several iterations, a sufficiently diverse dataset was obtained.

A separate validation dataset was then constructed to assess model performance. Seven representative initial structures were selected: FCC Ni, BCC Cr, HCP Co, equiatomic binary FCC alloys (CrCo, CrNi, CoNi), and ternary FCC CrCoNi. These systems were simulated in the isothermal–isobaric (NPT) ensemble over a temperature range of 5–3000 K for 2 ns. The extended simulation time ensures both model stability and thorough exploration of the configuration space. From each trajectory, 12 configurations were uniformly sampled, resulting in a validation dataset comprising 84 structures (Fig.~\ref{fig:dataset}). The active learning iterations were terminated when the RMSEs of energy, force, and stress for the validation dataset became comparable to those of the training dataset; otherwise, the validation structures were incorporated into the training set and the process was repeated. As summarized in Table~\ref{tab:error}, the RMSEs for the training and validation datasets are in close agreement. The final NEP dataset contains 3030 structures. The model was trained with optimized hyperparameters using eight Tesla V100 GPUs, requiring 57 hours and approximately 400,000 generations to achieve convergence.

\subsubsection{Hybrid MD/MC simulation}

The MD/MC simulations shown in Fig.~\ref{fig:mcmd} were initialized from chemically random configurations and performed in the NPT ensemble using a Nos\'{e}–Hoover thermostat and barostat to maintain zero pressure at temperatures of 500, 800, and 1200 K. Each system was equilibrated for 5000 steps in a 3$\times$3$\times$3 conventional FCC supercell containing 108 atoms. Subsequently, hybrid MD/MC simulations were carried out for an additional 5000 steps, during which atomic species were randomly swapped according to the Metropolis criterion within the NPT ensemble, following a procedure similar to previous DFT-based MC simulations~\cite{cao2025capturing}. The Warren–Cowley parameters were calculated by averaging over the final 100 steps of each simulation. For each interatomic potential, a total of 20 independent simulations were performed. All simulations were performed using the LAMMPS code~\cite{lammps} with a timestep of 1 fs. Atomic configurations were visualized using OVITO~\cite{ovito}, and all data pre- and post-processing were carried out using the self-developed open-source code MDAPY~\cite{mdapy}.

The stacking fault energy (SFE) shown in Fig.~\ref{fig:sfe_dislo}a was calculated using a supercell containing 216 atoms, with orthogonal axes aligned along the [112], [1$\bar{1}$0], and [11$\bar{1}$] directions of the FCC lattice. The SFE was obtained from the energy difference before and after displacing the upper half of the crystal along the z-direction by a distance of $a/\sqrt{6}$, where $a$ is the lattice constant. For each interatomic potential, the SFE was averaged over 150 structures with random elemental distributions. To evaluate the effect of short-range order (SRO), each of these 150 structures was further subjected to 5000 MC swap steps at 0 K, after which the SFE of the ordered configurations was calculated~(Fig.~\ref{fig:sfe_dislo}b). 

For the ternary phase diagram (Fig.~\ref{fig:tenary}), the composition of each element was varied from 0 to 100 at.\% in increments of 5 at.\%. For each composition, a 3$\times$3$\times$3 FCC supercell (108 atoms) and a 4$\times$4$\times$4 BCC supercell (128 atoms) were constructed and subjected to cell optimization and energy minimization using various interatomic potentials. The properties at each composition were averaged over five independent structures to obtain $\Delta_{\mathrm{FCC-BCC}}$ and the lattice constant $a$. Subsequently, SFE and shear modulus ($G$) were calculated only for compositions where the FCC phase is more stable than the BCC phase, using the optimized lattice constant. For each selected composition, 150 independent structures were generated for SFE calculations, while the shear modulus was evaluated using a single 3$\times$3$\times$3 FCC supercell (108 atoms).

\subsubsection{Dislocation dissociation}

To investigate dislocation dissociation (Fig.~\ref{fig:sfe_dislo}c-f), a Ni bulk system was constructed with crystallographic orientations along [110], [$\bar{1}$11], and [112], with dimensions of approximately 200$\times$152$\times$22 \AA. Two edge dislocations were introduced, with the dislocation line along the z-direction and the Burgers vector along the x-direction. This dipole configuration ensures a zero net Burgers vector, enabling fully periodic boundary conditions. A similar approach was employed to construct screw dislocation models using a supercell oriented along [1$\bar{1}$2], [$\bar{1}$11], and [110], with dimensions of approximately 258$\times$244$\times$25 \AA. In this case, four screw dislocations were introduced to form a quadrupole configuration, ensuring periodicity in all directions.

For the dipole edge configuration, the two dislocations were positioned at (x, y) coordinates of (0.25, 0.25) and (0.75, 0.75). For the quadrupole screw configuration, four dislocations were placed at (x, y) coordinates of (0.25, 0.25), (0.75, 0.25), (0.25, 0.75), and (0.75, 0.75). The maximum stacking fault widths (SFWs) observed in Fig.~\ref{fig:sfe_dislo} are approximately 65 \AA~for edge dislocations and 38 \AA~for screw dislocations in the CrCoNi alloy. These values are well below half of the box length in the edge dislocation model ($\sim$100 \AA) and one quarter in the screw dislocation model ($\sim$64.5 \AA) along the slip (x) direction, ensuring negligible interactions between dissociated dislocations and their periodic images.

All dislocation structures were generated using ATOMSK~\cite{atomsk}. The same dislocation models were further used to construct ten equiatomic CrCoNi samples with different elemental distributions. All systems were subjected to full cell optimization and energy minimization at 0 K using different interatomic potentials. The convergence criteria for energy and force are set to $1 \times 10^{-12}$~eV and $1 \times 10^{-12}$~eV/\AA, respectively. During relaxation, the dislocations dissociate along the x-direction into two partial dislocations separated by a stacking fault. The stacking fault width was averaged over the two (edge) or four (screw) dislocations in Ni, and further averaged over ten independent configurations for the CrCoNi alloy.

\subsubsection{Liquid state simulation}

The two-phase coexistence method was employed to determine the melting temperature ($\mathrm{T_m}$) of FCC Ni, BCC Cr, HCP Co, and the equiatomic FCC CrCoNi alloy. The supercell was constructed with a length of approximately 70 \AA~along the x-direction and ~36 \AA~along the other directions, containing between 8,000 and 9,400 atoms. Each supercell was divided along the x-direction into two regions: a solid region and a liquid region.

The system was first equilibrated at $\mathrm{T_{guess}} - 1000$ K for 20 ps, where $\mathrm{T_{guess}}$ denotes the initial estimate of the melting temperature. Then, the solid region was fixed, and the liquid region was directly set to $2\mathrm{T_{guess}}$ K and maintained for 40 ps using a Langevin thermostat to ensure complete melting. The system was subsequently evolved in the isobaric–isoenthalpic (NPH) ensemble, maintaining zero pressure, for 400 ps. If solid–liquid coexistence was sustained over this timescale, the melting temperature $\mathrm{T_m}$ was determined by averaging the temperature over the final 100 ps (Fig.~\ref{fig:liquid}a,b). If the system fully solidified, $\mathrm{T_{guess}}$ was increased; if it completely melted, $\mathrm{T_{guess}}$ was decreased. This procedure was iteratively repeated until the correct $\mathrm{T_m}$ was obtained.

For the radial distribution function shown in Fig.~\ref{fig:liquid}d, a 10$\times$10$\times$10 FCC supercell containing 4000 atoms was equilibrated in the NPT ensemble at 2684 K and zero hydrostatic pressure for 20 ps. The element-resolved partial RDFs were computed using configurations collected over the final 10 ps.

\section{Data availability}

All datasets and the trained NEP model will be made publicly available in a GitHub repository upon acceptance of the manuscript.

\section{Code availability}

The source code for MDAPY is available at the Github repository: https://github.com/mushroomfire/mdapy.
The source code for LAMMPS is available at the Github repository: https://github.com/lammps/lammps.
The source code for GPUMD is available at the Github repository: https://github.com/brucefan1983/GPUMD. 

\begin{acknowledgments}
Y.-C.~W., T.~M. and M.~A. acknowledge the support from FinnCERES flagship (grant no.~151830423), Business Finland (grant nos.~211835, 211909, and 211989), the Research Council of Finland (grant no.~13361245), and the Future Makers program. 
M.~A. acknowledges support from the Academy of Finland Center of Excellence program (program nos. 278367 and 317464), as well as the Finnish Cultural Foundation.
A.~E. acknowledges support from the European Union Horizon 2020 research and innovation program under Grant Agreement No.~857470, from the European Regional Development Fund under the Foundation for Polish Science International Research Agenda PLUS program (Grant No.~MAB PLUS/2018/8), and from the initiative of the Ministry of Science and Higher Education “Support for the activities of Centers of Excellence established in Poland under the Horizon 2020 program” (Agreement No.~MEiN/2023/DIR/3795).
The authors acknowledge the computational resources provided by the Aalto University School of Science “Science-IT” project, as well as by CSC (Finland) \textit{via} the project 2015437.
\end{acknowledgments}

\bibliography{myref}

@article{cao2025capturing,
	title = {Capturing short-range order in high-entropy alloys with machine learning potentials},
	volume = {11},
	issn = {2057-3960},
	url = {https://www.nature.com/articles/s41524-025-01722-2},
	doi = {10.1038/s41524-025-01722-2},
	number = {1},
	urldate = {2025-09-17},
	journal = {npj Comput Mater},
	author = {Cao, Yifan and Sheriff, Killian and Freitas, Rodrigo},
	month = aug,
	year = {2025},
	pages = {268},
}

@misc{nitol2025achieving,
	title = {Achieving {DFT} accuracy in short range ordering and stacking fault energy using moment tensor potential for {CoCrFeNi} and {CoCrNi}},
	url = {http://arxiv.org/abs/2509.11231},
	doi = {10.48550/arXiv.2509.11231},
	urldate = {2025-09-18},
	publisher = {arXiv},
	author = {Nitol, Mashroor S. and Tamm, Artur and Mubassira, Subah and Xu, Shuozhi and Fensin, Saryu J.},
	month = sep,
	year = {2025},
	note = {arXiv:2509.11231 [cond-mat]},
}

@article{li2019strengthening,
	title = {Strengthening in multi-principal element alloys with local-chemical-order roughened dislocation pathways},
	volume = {10},
	issn = {2041-1723},
	url = {https://www.nature.com/articles/s41467-019-11464-7},
	doi = {10.1038/s41467-019-11464-7},
	number = {1},
	urldate = {2025-09-16},
	journal = {Nat Commun},
	author = {Li, Qing-Jie and Sheng, Howard and Ma, Evan},
	month = aug,
	year = {2019},
	pages = {3563},
}

@article{farkas2020model,
	title = {Model interatomic potentials for {Fe}–{Ni}–{Cr}–{Co}–{Al} high-entropy alloys},
	volume = {35},
	issn = {0884-2914, 2044-5326},
	url = {http://link.springer.com/10.1557/jmr.2020.294},
	doi = {10.1557/jmr.2020.294},
	number = {22},
	urldate = {2026-01-08},
	journal = {J. Mater. Res.},
	author = {Farkas, Diana and Caro, Alfredo},
	month = nov,
	year = {2020},
	pages = {3031--3040},
}

@article{laplanche2020processing,
	title = {Processing of a single-crystalline {CrCoNi} medium-entropy alloy and evolution of its thermal expansion and elastic stiffness coefficients with temperature},
	volume = {177},
	issn = {13596462},
	url = {https://linkinghub.elsevier.com/retrieve/pii/S1359646219305548},
	doi = {10.1016/j.scriptamat.2019.09.020},
	urldate = {2025-10-15},
	journal = {Scripta Materialia},
	author = {Laplanche, G. and Schneider, M. and Scholz, F. and Frenzel, J. and Eggeler, G. and Schreuer, J.},
	month = mar,
	year = {2020},
	pages = {44--48},
}

@article{ge2018effect,
	title = {Effect of alloying on the thermal-elastic properties of 3d high-entropy alloys},
	volume = {210},
	issn = {02540584},
	url = {https://linkinghub.elsevier.com/retrieve/pii/S0254058417308350},
	doi = {10.1016/j.matchemphys.2017.10.046},
	urldate = {2025-10-15},
	journal = {Materials Chemistry and Physics},
	author = {Ge, Huijuan and Song, Hongquan and Shen, Jiang and Tian, Fuyang},
	month = may,
	year = {2018},
	pages = {320--326},
}

@article{ding2018tunable,
author = {Jun Ding  and Qin Yu  and Mark Asta  and Robert O. Ritchie },
title = {Tunable stacking fault energies by tailoring local chemical order in CrCoNi medium-entropy alloys},
journal = {Proceedings of the National Academy of Sciences},
volume = {115},
number = {36},
pages = {8919-8924},
year = {2018},
doi = {10.1073/pnas.1808660115},
URL = {https://www.pnas.org/doi/abs/10.1073/pnas.1808660115},

}

@article{tamm2015atomic,
title = {Atomic-scale properties of Ni-based FCC ternary, and quaternary alloys},
journal = {Acta Materialia},
volume = {99},
pages = {307-312},
year = {2015},
issn = {1359-6454},
doi = {https://doi.org/10.1016/j.actamat.2015.08.015},
url = {https://www.sciencedirect.com/science/article/pii/S1359645415005790},
author = {Artur Tamm and Alvo Aabloo and Mattias Klintenberg and Malcolm Stocks and Alfredo Caro},
}

@article{kong2015phase,
	title = {Phase diagram, mechanical and thermodynamics properties of metallic {Co} under high temperature and high pressure},
	volume = {104},
	issn = {09270256},
	url = {https://linkinghub.elsevier.com/retrieve/pii/S0927025615002190},
	doi = {10.1016/j.commatsci.2015.03.046},
	urldate = {2026-02-02},
	journal = {Computational Materials Science},
	author = {Kong, Bo and Zeng, Ti-Xian and Xu, Hong-Bin and Chen, De-liang and Zhou, Zhu-Wen and Fu, Zhi-Jian},
	month = jun,
	year = {2015},
	pages = {130--137},
}

@article{antonangeli2004elasticity,
	title = {Elasticity of {Cobalt} at {High} {Pressure} {Studied} by {Inelastic} {X}-{Ray} {Scattering}},
	volume = {93},
	copyright = {http://link.aps.org/licenses/aps-default-license},
	issn = {0031-9007, 1079-7114},
	url = {https://link.aps.org/doi/10.1103/PhysRevLett.93.215505},
	doi = {10.1103/PhysRevLett.93.215505},
	number = {21},
	urldate = {2026-02-02},
	journal = {Phys. Rev. Lett.},
	author = {Antonangeli, D. and Krisch, M. and Fiquet, G. and Farber, D. L. and Aracne, C. M. and Badro, J. and Occelli, F. and Requardt, H.},
	month = nov,
	year = {2004},
	pages = {215505},
}

@article{dewaele2008compression,
	title = {Compression curves of transition metals in the {Mbar} range: {Experiments} and projector augmented-wave calculations},
	volume = {78},
	copyright = {http://link.aps.org/licenses/aps-default-license},
	issn = {1098-0121, 1550-235X},
	shorttitle = {Compression curves of transition metals in the {Mbar} range},
	url = {https://link.aps.org/doi/10.1103/PhysRevB.78.104102},
	doi = {10.1103/PhysRevB.78.104102},
	number = {10},
	urldate = {2026-02-02},
	journal = {Phys. Rev. B},
	author = {Dewaele, Agn\`{e}s and Torrent, Marc and Loubeyre, Paul and Mezouar, Mohamed},
	month = sep,
	year = {2008},
	pages = {104102},

}

@article{ioroi2022melting,
	title = {Melting {Point} of {Pure} {Cr} and {Phase} {Equilibria} in the {Cr}-{Si} {Binary} {System}},
	volume = {43},
	issn = {1547-7037, 1863-7345},
	url = {https://link.springer.com/10.1007/s11669-022-00954-9},
	doi = {10.1007/s11669-022-00954-9},
	number = {2},
	urldate = {2026-02-03},
	journal = {J. Phase Equilib. Diffus.},
	author = {Ioroi, Kazushige and Aono, Yuki and Xu, Xiao and Omori, Toshihiro and Kainuma, Ryosuke},
	month = apr,
	year = {2022},
	pages = {229--242},
}

@article{bideault2024polyvalent,
	title = {Polyvalent machine-learned potential for cobalt: {From} bulk to nanoparticles},
	volume = {8},
	issn = {2475-9953},
	shorttitle = {Polyvalent machine-learned potential for cobalt},
	url = {https://link.aps.org/doi/10.1103/PhysRevMaterials.8.123803},
	doi = {10.1103/PhysRevMaterials.8.123803},
	number = {12},
	urldate = {2026-02-03},
	journal = {Phys. Rev. Materials},
	author = {Bideault, Marthe and Creuze, J\'{e}r\^{o}me and Asahi, Ryoji and Wimmer, Erich},
	month = dec,
	year = {2024},
	pages = {123803},
}

@article{razumovskiy2011firstprinciples,
	title = {First-principles study of elastic properties of {Cr}- and {Fe}-rich {Fe}-{Cr} alloys},
	volume = {84},
	copyright = {http://link.aps.org/licenses/aps-default-license},
	issn = {1098-0121, 1550-235X},
	url = {https://link.aps.org/doi/10.1103/PhysRevB.84.024106},
	doi = {10.1103/PhysRevB.84.024106},
	number = {2},
	urldate = {2026-02-05},
	journal = {Phys. Rev. B},
	author = {Razumovskiy, Vsevolod I. and Ruban, Andrei V. and Korzhavyi, Pavel A.},
	month = jul,
	year = {2011},
	pages = {024106},
}

@article{gao2013phase,
	title = {Phase stability and elastic properties of {Cr}–{V} alloys},
	volume = {25},
	copyright = {http://iopscience.iop.org/info/page/text-and-data-mining},
	issn = {0953-8984, 1361-648X},
	url = {https://iopscience.iop.org/article/10.1088/0953-8984/25/7/075402},
	doi = {10.1088/0953-8984/25/7/075402},
	number = {7},
	urldate = {2026-02-05},
	journal = {J. Phys.: Condens. Matter},
	author = {Gao, M C and Suzuki, Y and Schweiger, H and Do\u{g}an, \"{O} N and Hawk, J and Widom, M},
	month = feb,
	year = {2013},
	pages = {075402},
}

@article{soderlind1994crystal,
	title = {Crystal structure and elastic-constant anomalies in the magnetic 3 \textit{d} transition metals},
	volume = {50},
	copyright = {http://link.aps.org/licenses/aps-default-license},
	issn = {0163-1829, 1095-3795},
	url = {https://link.aps.org/doi/10.1103/PhysRevB.50.5918},
	doi = {10.1103/PhysRevB.50.5918},
	number = {9},
	urldate = {2026-02-05},
	journal = {Phys. Rev. B},
	author = {S\"{o}derlind, P. and Ahuja, R. and Eriksson, O. and Wills, J. M. and Johansson, B.},
	month = sep,
	year = {1994},
	pages = {5918--5927},
}

@article{gong2024accurate,
	title = {An accurate and transferable machine learning interatomic potential for nickel},
	volume = {5},
	issn = {2662-4443},
	url = {https://www.nature.com/articles/s43246-024-00603-3},
	doi = {10.1038/s43246-024-00603-3},
	number = {1},
	urldate = {2026-02-05},
	journal = {Commun Mater},
	author = {Gong, Xiaoguo and Li, Zhuoyuan and Pattamatta, A. S. L. Subrahmanyam and Wen, Tongqi and Srolovitz, David J.},
	month = aug,
	year = {2024},
	pages = {157},

}

@article{kanhe2016investigation,
	title = {Investigation of structural and magnetic properties of thermal plasma-synthesized {Fe1}−{xNi} alloy nanoparticles},
	volume = {663},
	issn = {09258388},
	url = {https://linkinghub.elsevier.com/retrieve/pii/S0925838815317473},
	doi = {10.1016/j.jallcom.2015.11.190},
	urldate = {2026-02-05},
	journal = {Journal of Alloys and Compounds},
	author = {Kanhe, Nilesh S. and Kumar, Amit and Yusuf, S.M. and Nawale, A.B. and Gaikwad, S.S. and Raut, Suyog A. and Bhoraskar, S.V. and Wu, Sheng Yun and Das, A.K. and Mathe, V.L.},
	month = apr,
	year = {2016},
	pages = {30--40},
}

@article{megchiche2006density,
	title = {Density functional calculations of the formation and migration enthalpies of monovacancies in {Ni}: {Comparison} of local and nonlocal approaches},
	volume = {74},
	copyright = {http://link.aps.org/licenses/aps-default-license},
	issn = {1098-0121, 1550-235X},
	shorttitle = {Density functional calculations of the formation and migration enthalpies of monovacancies in {Ni}},
	url = {https://link.aps.org/doi/10.1103/PhysRevB.74.064111},
	doi = {10.1103/PhysRevB.74.064111},
	number = {6},
	urldate = {2026-02-05},
	journal = {Phys. Rev. B},
	author = {Megchiche, El Hocine and P\'{e}rusin, Simon and Barthelat, Jean-Claude and Mijoule, Claude},
	month = aug,
	year = {2006},
	pages = {064111},
}

@article{birgeneau1964normal,
	title = {Normal {Modes} of {Vibration} in {Nickel}},
	volume = {136},
	copyright = {http://link.aps.org/licenses/aps-default-license},
	issn = {0031-899X},
	url = {https://link.aps.org/doi/10.1103/PhysRev.136.A1359},
	doi = {10.1103/PhysRev.136.A1359},
	number = {5A},
	urldate = {2026-02-05},
	journal = {Phys. Rev.},
	author = {Birgeneau, R. J. and Cordes, J. and Dolling, G. and Woods, A. D. B.},
	month = nov,
	year = {1964},
	pages = {A1359--A1365},
}

@article{shaw1971investigation,
	title = {Investigation of the {Phonon} {Dispersion} {Relations} of {Chromium} by {Inelastic} {Neutron} {Scattering}},
	volume = {4},
	copyright = {http://link.aps.org/licenses/aps-default-license},
	issn = {0556-2805},
	url = {https://link.aps.org/doi/10.1103/PhysRevB.4.969},
	doi = {10.1103/PhysRevB.4.969},
	number = {3},
	urldate = {2026-02-05},
	journal = {Phys. Rev. B},
	author = {Shaw, W. M. and Muhlestein, L. D.},
	month = aug,
	year = {1971},
	pages = {969--973},
}

@article{carter1977the,
author = {C. B. Carter and S. M. Holmes},
title = {The stacking-fault energy of nickel},
journal = {The Philosophical Magazine: A Journal of Theoretical Experimental and Applied Physics},
volume = {35},
number = {5},
pages = {1161--1172},
year = {1977},
publisher = {Taylor \& Francis},
doi = {10.1080/14786437708232942},
}

@article{choi2018understanding,
  title={Understanding the physical metallurgy of the CoCrFeMnNi high-entropy alloy: an atomistic simulation study},
  author={Choi, Won-Mi and Jo, Yong Hee and Sohn, Seok Su and Lee, Sunghak and Lee, Byeong-Joo},
  journal={npj Computational Materials},
  volume={4},
  number={1},
  pages={1},
  year={2018},
  publisher={Nature Publishing Group UK London}
}

@article{wakaba1982lattice,
  title = {Lattice dynamic of Ti, Co, Tc, and other hcp transition metals},
  author = {Wakabayashi, N. and Scherm, R. H. and Smith, H. G.},
  journal = {Phys. Rev. B},
  volume = {25},
  issue = {8},
  pages = {5122--5132},
  numpages = {0},
  year = {1982},
  month = {Apr},
  publisher = {American Physical Society},
  doi = {10.1103/PhysRevB.25.5122},
  url = {https://link.aps.org/doi/10.1103/PhysRevB.25.5122}
}

@article{wu2014temperature,
title = {Temperature dependence of the mechanical properties of equiatomic solid solution alloys with face-centered cubic crystal structures},
journal = {Acta Materialia},
volume = {81},
pages = {428-441},
year = {2014},
issn = {1359-6454},
doi = {https://doi.org/10.1016/j.actamat.2014.08.026},
url = {https://www.sciencedirect.com/science/article/pii/S1359645414006272},
author = {Z. Wu and H. Bei and G.M. Pharr and E.P. George},
}

@article{zhu2023effects,
title = {Effects of short range ordering on the generalized stacking fault energy and deformation mechanisms in FCC multiprincipal element alloys},
journal = {Acta Materialia},
volume = {259},
pages = {119230},
year = {2023},
issn = {1359-6454},
doi = {https://doi.org/10.1016/j.actamat.2023.119230},
url = {https://www.sciencedirect.com/science/article/pii/S1359645423005608},
author = {Lingyu Zhu and Zhaoxuan Wu},
}

@article{laplanche2017reasons,
title = {Reasons for the superior mechanical properties of medium-entropy CrCoNi compared to high-entropy CrMnFeCoNi},
journal = {Acta Materialia},
volume = {128},
pages = {292-303},
year = {2017},
issn = {1359-6454},
doi = {https://doi.org/10.1016/j.actamat.2017.02.036},
url = {https://www.sciencedirect.com/science/article/pii/S135964541730126X},
author = {G. Laplanche and A. Kostka and C. Reinhart and J. Hunfeld and G. Eggeler and E.P. George},
}

@article{liu2018stacking,
title = {Stacking fault energy of face-centered-cubic high entropy alloys},
journal = {Intermetallics},
volume = {93},
pages = {269-273},
year = {2018},
issn = {0966-9795},
doi = {https://doi.org/10.1016/j.intermet.2017.10.004},
url = {https://www.sciencedirect.com/science/article/pii/S0966979517307379},
author = {S.F. Liu and Y. Wu and H.T. Wang and J.Y. He and J.B. Liu and C.X. Chen and X.J. Liu and H. Wang and Z.P. Lu},
}

@article{deng2021enhancement,
title = {Enhancement of strength and ductility in non-equiatomic CoCrNi medium-entropy alloy at room temperature via transformation-induced plasticity},
journal = {Materials Science and Engineering: A},
volume = {804},
pages = {140516},
year = {2021},
issn = {0921-5093},
doi = {https://doi.org/10.1016/j.msea.2020.140516},
url = {https://www.sciencedirect.com/science/article/pii/S0921509320315793},
author = {H.W. Deng and M.M. Wang and Z.M. Xie and T. Zhang and X.P. Wang and Q.F. Fang and Y. Xiong},
}

@article{coury2019design,
title = {Design and in-situ characterization of a strong and ductile co-rich multicomponent alloy with transformation induced plasticity},
journal = {Scripta Materialia},
volume = {173},
pages = {70-74},
year = {2019},
issn = {1359-6462},
doi = {https://doi.org/10.1016/j.scriptamat.2019.07.045},
url = {https://www.sciencedirect.com/science/article/pii/S1359646219304634},
author = {Francisco Gil Coury and Diego Santana and Yaofeng Guo and John Copley and Lucas Otani and Solange Fonseca and Guilherme Zepon and Claudio Kiminami and Michael Kaufman and Amy Clarke},
}

@article{coury2018highthrou,
  title={High throughput discovery and design of strong multicomponent metallic solid solutions},
  author={Coury, Francisco G and Clarke, Kester D and Kiminami, Claudio S and Kaufman, Michael J and Clarke, Amy J},
  journal={Scientific reports},
  volume={8},
  number={1},
  pages={8600},
  year={2018},
  publisher={Nature Publishing Group UK London}
}

@article{yang2022theoretical,
title = {Theoretical and experimental study of phase transformation and twinning behavior in metastable high-entropy alloys},
journal = {Journal of Materials Science \& Technology},
volume = {99},
pages = {161-168},
year = {2022},
issn = {1005-0302},
doi = {https://doi.org/10.1016/j.jmst.2021.05.037},
url = {https://www.sciencedirect.com/science/article/pii/S1005030221005776},
author = {Zhibiao Yang and Song Lu and Yanzhong Tian and Zijian Gu and Jian Sun and Levente Vitos},
}

@article{shuhei2020effectof,
  title={Effect of Cobalt-Content on Mechanical Properties of Non-Equiatomic Co–Cr–Ni Medium Entropy Alloys},
  author={Shuhei Yoshida and Takuto Ikeuchi and Yu Bai and Nobuhiro Tsuji},
  journal={MATERIALS TRANSACTIONS},
  volume={61},
  number={4},
  pages={587-595},
  year={2020},
  doi={10.2320/matertrans.MT-MK2019004}
}

@article{huang2022advanced,
title = {Advanced mechanical properties obtained via accurately tailoring stacking fault energy in Co-rich and Ni-depleted CoxCr33Ni67-x medium-entropy alloys},
journal = {Scripta Materialia},
volume = {207},
pages = {114269},
year = {2022},
issn = {1359-6462},
doi = {https://doi.org/10.1016/j.scriptamat.2021.114269},
url = {https://www.sciencedirect.com/science/article/pii/S1359646221005492},
author = {Dong Huang and Yanxin Zhuang and Chunhui Wang},
}

@article{gustavo2021hallpetch,
    author = {Bertoli, Gustavo and Otani, Lucas B. and Clarke, Amy J. and Kiminami, Claudio S. and Coury, Francisco G.},
    title = {Hall–Petch and grain growth kinetics of the low stacking fault energy TRIP Cr40Co40Ni20 multi-principal element alloy},
    journal = {Applied Physics Letters},
    volume = {119},
    number = {6},
    pages = {061903},
    year = {2021},
    month = {08},
    issn = {0003-6951},
    doi = {10.1063/5.0057888},
    url = {https://doi.org/10.1063/5.0057888},

}

@article{yoshida2019deformation,
doi = {10.1088/1757-899X/580/1/012053},
url = {https://doi.org/10.1088/1757-899X/580/1/012053},
year = {2019},
month = {aug},
publisher = {IOP Publishing},
volume = {580},
number = {1},
pages = {012053},
author = {Yoshida, Shuhei and Ikeuchi, Takuto and Bai, Yu and Shibata, Akinobu and Hansen, Niels and Huang, Xiaoxu and Tsuji, Nobuhiro},
title = {Deformation microstructures and strength of face-centered cubic high/medium entropy alloys},
journal = {IOP Conference Series: Materials Science and Engineering},
}

@article{madhavan2014texture,
title = {Texture transition in cold-rolled nickel–40wt.\% cobalt alloy},
journal = {Acta Materialia},
volume = {74},
pages = {151-164},
year = {2014},
issn = {1359-6454},
doi = {https://doi.org/10.1016/j.actamat.2014.03.066},
url = {https://www.sciencedirect.com/science/article/pii/S1359645414002341},
author = {R. Madhavan and R.K. Ray and S. Suwas},
}

@article{tan2019dislocation,
  title = {Dislocation core structures in Ni-based superalloys computed using a density functional theory based flexible boundary condition approach},
  author = {Tan, Anne Marie Z. and Woodward, Christopher and Trinkle, Dallas R.},
  journal = {Phys. Rev. Mater.},
  volume = {3},
  issue = {3},
  pages = {033609},
  numpages = {8},
  year = {2019},
  month = {Mar},
  publisher = {American Physical Society},
  doi = {10.1103/PhysRevMaterials.3.033609},
  url = {https://link.aps.org/doi/10.1103/PhysRevMaterials.3.033609}
}

@article{korhonen1995vacancy,
  title = {Vacancy-formation energies for fcc and bcc transition metals},
  author = {Korhonen, T. and Puska, M. J. and Nieminen, R. M.},
  journal = {Phys. Rev. B},
  volume = {51},
  issue = {15},
  pages = {9526--9532},
  numpages = {0},
  year = {1995},
  month = {Apr},
  publisher = {American Physical Society},
  doi = {10.1103/PhysRevB.51.9526},
  url = {https://link.aps.org/doi/10.1103/PhysRevB.51.9526}
}

@article{george2019high,
  title={High-entropy alloys},
  author={George, Easo P and Raabe, Dierk and Ritchie, Robert O},
  journal={Nature reviews materials},
  volume={4},
  number={8},
  pages={515--534},
  year={2019},
  publisher={Nature Publishing Group UK London}
}

@article{li2019mechanical,
title = {Mechanical properties of high-entropy alloys with emphasis on face-centered cubic alloys},
journal = {Progress in Materials Science},
volume = {102},
pages = {296-345},
year = {2019},
issn = {0079-6425},
doi = {https://doi.org/10.1016/j.pmatsci.2018.12.003},
url = {https://www.sciencedirect.com/science/article/pii/S0079642518301178},
author = {Zezhou Li and Shiteng Zhao and Robert O. Ritchie and Marc A. Meyers},
}

@article{granberg2016mechanism,
  title = {Mechanism of Radiation Damage Reduction in Equiatomic Multicomponent Single Phase Alloys},
  author = {Granberg, F. and Nordlund, K. and Ullah, Mohammad W. and Jin, K. and Lu, C. and Bei, H. and Wang, L. M. and Djurabekova, F. and Weber, W. J. and Zhang, Y.},
  journal = {Phys. Rev. Lett.},
  volume = {116},
  issue = {13},
  pages = {135504},
  numpages = {8},
  year = {2016},
  month = {Apr},
  publisher = {American Physical Society},
  doi = {10.1103/PhysRevLett.116.135504},
  url = {https://link.aps.org/doi/10.1103/PhysRevLett.116.135504}
}

@article{zhao2023deformation,
author = {Shiteng Zhao  and Sheng Yin  and Xiao Liang  and Fuhua Cao  and Qin Yu  and Ruopeng Zhang  and Lanhong Dai  and Carlos J. Ruestes  and Robert O. Ritchie  and Andrew M. Minor },
title = {Deformation and failure of the CrCoNi medium-entropy alloy subjected to extreme shock loading},
journal = {Science Advances},
volume = {9},
number = {18},
pages = {eadf8602},
year = {2023},
doi = {10.1126/sciadv.adf8602},
URL = {https://www.science.org/doi/abs/10.1126/sciadv.adf8602},
}

@article{yang2019highimpact,
title = {High impact toughness of CrCoNi medium-entropy alloy at liquid-helium temperature},
journal = {Scripta Materialia},
volume = {172},
pages = {66-71},
year = {2019},
issn = {1359-6462},
doi = {https://doi.org/10.1016/j.scriptamat.2019.07.010},
url = {https://www.sciencedirect.com/science/article/pii/S1359646219304087},
author = {Muxin Yang and Lingling Zhou and Chang Wang and Ping Jiang and Fuping Yuan and Evan Ma and Xiaolei Wu},
}

@article{liu2022exceptional,
author = {Dong Liu  and Qin Yu  and Saurabh Kabra  and Ming Jiang  and Paul Forna-Kreutzer  and Ruopeng Zhang  and Madelyn Payne  and Flynn Walsh  and Bernd Gludovatz  and Mark Asta  and Andrew M. Minor  and Easo P. George  and Robert O. Ritchie },
title = {Exceptional fracture toughness of CrCoNi-based medium- and high-entropy alloys at 20 kelvin},
journal = {Science},
volume = {378},
number = {6623},
pages = {978-983},
year = {2022},
doi = {10.1126/science.abp8070},
URL = {https://www.science.org/doi/abs/10.1126/science.abp8070},
}

@article{slone2018influence,
title = {Influence of deformation induced nanoscale twinning and FCC-HCP transformation on hardening and texture development in medium-entropy CrCoNi alloy},
journal = {Acta Materialia},
volume = {158},
pages = {38-52},
year = {2018},
issn = {1359-6454},
doi = {https://doi.org/10.1016/j.actamat.2018.07.028},
url = {https://www.sciencedirect.com/science/article/pii/S1359645418305561},
author = {C.E. Slone and S. Chakraborty and J. Miao and E.P. George and M.J. Mills and S.R. Niezgoda},
}

@article{coury2021multi,
title = {Multi-principal element alloys from the CrCoNi family: outlook and perspectives},
journal = {Journal of Materials Research and Technology},
volume = {15},
pages = {3461-3480},
year = {2021},
issn = {2238-7854},
doi = {https://doi.org/10.1016/j.jmrt.2021.09.095},
url = {https://www.sciencedirect.com/science/article/pii/S223878542101084X},
author = {Francisco G. Coury and Guilherme Zepon and Claudemiro Bolfarini},
}

@article{zhang2020short,
  title={Short-range order and its impact on the CrCoNi medium-entropy alloy},
  author={Zhang, Ruopeng and Zhao, Shiteng and Ding, Jun and Chong, Yan and Jia, Tao and Ophus, Colin and Asta, Mark and Ritchie, Robert O and Minor, Andrew M},
  journal={Nature},
  volume={581},
  number={7808},
  pages={283--287},
  year={2020},
  publisher={Nature Publishing Group UK London}
}

@article{shih2021stacking,
  title={Stacking fault energy in concentrated alloys},
  author={Shih, Mulaine and Miao, Jiashi and Mills, Michael and Ghazisaeidi, Maryam},
  journal={Nature communications},
  volume={12},
  number={1},
  pages={3590},
  year={2021},
  publisher={Nature Publishing Group UK London}
}

@article{rasooli2026searching,
title = {Searching for evidence of strengthening from short-range order in the CrCoNi medium entropy alloy},
journal = {Scripta Materialia},
volume = {271},
pages = {116997},
year = {2026},
issn = {1359-6462},
doi = {https://doi.org/10.1016/j.scriptamat.2025.116997},
url = {https://www.sciencedirect.com/science/article/pii/S1359646225004592},
author = {Novin Rasooli and Matthew Daly},
}

@article{li2023evolution,
title = {Evolution of short-range order and its effects on the plastic deformation behavior of single crystals of the equiatomic Cr-Co-Ni medium-entropy alloy},
journal = {Acta Materialia},
volume = {243},
pages = {118537},
year = {2023},
issn = {1359-6454},
doi = {https://doi.org/10.1016/j.actamat.2022.118537},
url = {https://www.sciencedirect.com/science/article/pii/S1359645422009144},
author = {Le Li and Zhenghao Chen and Shogo Kuroiwa and Mitsuhiro Ito and Koretaka Yuge and Kyosuke Kishida and Hisanori Tanimoto and Yue Yu and Haruyuki Inui and Easo P. George},
}

@article{lee2000secondn,
  title = {Second nearest-neighbor modified embedded-atom-method potential},
  author = {Lee, Byeong-Joo and Baskes, M. I.},
  journal = {Phys. Rev. B},
  volume = {62},
  issue = {13},
  pages = {8564--8567},
  numpages = {0},
  year = {2000},
  month = {Oct},
  publisher = {American Physical Society},
  doi = {10.1103/PhysRevB.62.8564},
  url = {https://link.aps.org/doi/10.1103/PhysRevB.62.8564}
}

@article{daw1984embedded,
  title = {Embedded-atom method: Derivation and application to impurities, surfaces, and other defects in metals},
  author = {Daw, Murray S. and Baskes, M. I.},
  journal = {Phys. Rev. B},
  volume = {29},
  issue = {12},
  pages = {6443--6453},
  numpages = {0},
  year = {1984},
  month = {Jun},
  publisher = {American Physical Society},
  doi = {10.1103/PhysRevB.29.6443},
  url = {https://link.aps.org/doi/10.1103/PhysRevB.29.6443}
}

@article{mishin2021machine,
title = {Machine-learning interatomic potentials for materials science},
journal = {Acta Materialia},
volume = {214},
pages = {116980},
year = {2021},
issn = {1359-6454},
doi = {https://doi.org/10.1016/j.actamat.2021.116980},
url = {https://www.sciencedirect.com/science/article/pii/S1359645421003608},
author = {Y. Mishin}
}

@article{novikov2022magnetic,
  title={Magnetic Moment Tensor Potentials for collinear spin-polarized materials reproduce different magnetic states of bcc Fe},
  author={Novikov, Ivan and Grabowski, Blazej and K\"{o}rmann, Fritz and Shapeev, Alexander},
  journal={npj Computational Materials},
  volume={8},
  number={1},
  pages={13},
  year={2022},
  publisher={Nature Publishing Group UK London}
}

@article{evgeny2023mlip3,
    author = {Podryabinkin, Evgeny and Garifullin, Kamil and Shapeev, Alexander and Novikov, Ivan},
    title = {MLIP-3: Active learning on atomic environments with moment tensor potentials},
    journal = {The Journal of Chemical Physics},
    volume = {159},
    number = {8},
    pages = {084112},
    year = {2023},
    month = {08},
    issn = {0021-9606},
    doi = {10.1063/5.0155887},
    url = {https://doi.org/10.1063/5.0155887},

}

@article{song2024general,
  title={General-purpose machine-learned potential for 16 elemental metals and their alloys},
  author={Song, Keke and Zhao, Rui and Liu, Jiahui and Wang, Yanzhou and Lindgren, Eric and Wang, Yong and Chen, Shunda and Xu, Ke and Liang, Ting and Ying, Penghua and others},
  journal={Nature Communications},
  volume={15},
  number={1},
  pages={10208},
  year={2024},
  publisher={Nature Publishing Group UK London}
}

@article{fan2021neuroevolution,
  title = {Neuroevolution machine learning potentials: Combining high accuracy and low cost in atomistic simulations and application to heat transport},
  author = {Fan, Zheyong and Zeng, Zezhu and Zhang, Cunzhi and Wang, Yanzhou and Song, Keke and Dong, Haikuan and Chen, Yue and Ala-Nissila, Tapio},
  journal = {Phys. Rev. B},
  volume = {104},
  issue = {10},
  pages = {104309},
  numpages = {15},
  year = {2021},
  month = {Sep},
  publisher = {American Physical Society},
  doi = {10.1103/PhysRevB.104.104309},
  url = {https://link.aps.org/doi/10.1103/PhysRevB.104.104309}
}

@article{zhang2024frontiers,
  title={Frontiers in high entropy alloys and high entropy functional materials},
  author={Zhang, Wen-Tao and Wang, Xue-Qian and Zhang, Feng-Qi and Cui, Xiao-Ya and Fan, Bing-Bing and Guo, Jia-Ming and Guo, Zhi-Min and Huang, Rui and Huang, Wen and Li, Xu-Bo and others},
  journal={Rare Metals},
  volume={43},
  number={10},
  pages={4639--4776},
  year={2024},
  publisher={Springer}
}

@article{du2022chemical,
title = {Chemical domain structure and its formation kinetics in CrCoNi medium-entropy alloy},
journal = {Acta Materialia},
volume = {240},
pages = {118314},
year = {2022},
issn = {1359-6454},
doi = {https://doi.org/10.1016/j.actamat.2022.118314},
url = {https://www.sciencedirect.com/science/article/pii/S1359645422006930},
author = {Jun-Ping Du and Peijun Yu and Shuhei Shinzato and Fan-Shun Meng and Yuji Sato and Yangen Li and Yiwen Fan and Shigenobu Ogata},
}

@article{hua2023revealing,
title = {Revealing the deformation mechanisms of 〈110〉 symmetric tilt grain boundaries in CoCrNi medium-entropy alloy},
journal = {International Journal of Plasticity},
volume = {171},
pages = {103832},
year = {2023},
issn = {0749-6419},
doi = {https://doi.org/10.1016/j.ijplas.2023.103832},
url = {https://www.sciencedirect.com/science/article/pii/S0749641923003169},
author = {Dongpeng Hua and Qing Zhou and Yeran Shi and Shuo Li and Ke Hua and Haifeng Wang and Suzhi Li and Weimin Liu},
}

@article{han2024ubiquitous,
  title={Ubiquitous short-range order in multi-principal element alloys},
  author={Han, Ying and Chen, Hangman and Sun, Yongwen and Liu, Jian and Wei, Shaolou and Xie, Bijun and Zhang, Zhiyu and Zhu, Yingxin and Li, Meng and Yang, Judith and others},
  journal={Nature Communications},
  volume={15},
  number={1},
  pages={6486},
  year={2024},
  publisher={Nature Publishing Group UK London}
}

@article{yan2023effectof,
title = {Effect of specimen size and crystallographic orientation on the nano/microscale mechanical properties and deformation behavior of CrCoNi medium-entropy alloy},
journal = {Materials \& Design},
volume = {235},
pages = {112387},
year = {2023},
issn = {0264-1275},
doi = {https://doi.org/10.1016/j.matdes.2023.112387},
url = {https://www.sciencedirect.com/science/article/pii/S026412752300802X},
author = {Shaohua Yan and Yuan Nie and Anna Paradowska},
}

@article{zhu2021unprecedented,
title = {Unprecedented combination of strength and ductility in laser welded NiCoCr medium entropy alloy joints},
journal = {Materials Science and Engineering: A},
volume = {803},
pages = {140501},
year = {2021},
issn = {0921-5093},
doi = {https://doi.org/10.1016/j.msea.2020.140501},
url = {https://www.sciencedirect.com/science/article/pii/S0921509320315641},
author = {Zhongyin Zhu and Shaohua Yan and Hui Chen and Guoqing Gou},
}

@article{utt2022origin,
  title={The origin of jerky dislocation motion in high-entropy alloys},
  author={Utt, Daniel and Lee, Subin and Xing, Yaolong and Jeong, Hyejin and Stukowski, Alexander and Oh, Sang Ho and Dehm, Gerhard and Albe, Karsten},
  journal={Nature communications},
  volume={13},
  number={1},
  pages={4777},
  year={2022},
  publisher={Nature Publishing Group UK London}
}

@article{amin2022edge,
  title = {Edge dislocations in multicomponent solid solution alloys: Beyond traditional elastic depinning},
  author = {Esfandiarpour, A. and Papanikolaou, S. and Alava, M.},
  journal = {Phys. Rev. Res.},
  volume = {4},
  issue = {2},
  pages = {L022043},
  numpages = {6},
  year = {2022},
  month = {May},
  publisher = {American Physical Society},
  doi = {10.1103/PhysRevResearch.4.L022043},
  url = {https://link.aps.org/doi/10.1103/PhysRevResearch.4.L022043}
}

@article{randle2006mechanisms,
title = {Mechanisms of grain boundary engineering},
journal = {Acta Materialia},
volume = {54},
number = {7},
pages = {1777-1783},
year = {2006},
issn = {1359-6454},
doi = {https://doi.org/10.1016/j.actamat.2005.11.046},
url = {https://www.sciencedirect.com/science/article/pii/S1359645406000073},
author = {Valerie Randle and Gregory Owen},
}

@article{ghosh2022short,
  title = {Short-range order and phase stability of CrCoNi explored with machine learning potentials},
  author = {Ghosh, Sheuly and Sotskov, Vadim and Shapeev, Alexander V. and Neugebauer, J\"{o}rg and K\"{o}rmann, Fritz},
  journal = {Phys. Rev. Mater.},
  volume = {6},
  issue = {11},
  pages = {113804},
  numpages = {10},
  year = {2022},
  month = {Nov},
  publisher = {American Physical Society},
  doi = {10.1103/PhysRevMaterials.6.113804},
  url = {https://link.aps.org/doi/10.1103/PhysRevMaterials.6.113804}
}

@article{yan2023design,
title = {Design and optimization of the composition and mechanical properties for non-equiatomic CoCrNi medium-entropy alloys},
journal = {Journal of Materials Science \& Technology},
volume = {139},
pages = {232-244},
year = {2023},
issn = {1005-0302},
doi = {https://doi.org/10.1016/j.jmst.2022.07.031},
url = {https://www.sciencedirect.com/science/article/pii/S1005030222006405},
author = {J.X. Yan and Z.J. Zhang and P. Zhang and J.H. Liu and H. Yu and Q.M. Hu and J.B. Yang and Z.F. Zhang},
}

@article{zhang2017origin,
  title={The origin of negative stacking fault energies and nano-twin formation in face-centered cubic high entropy alloys},
  author={Zhang, YH and Zhuang, Yu and Hu, Alice and Kai, Ji-Jung and Liu, Chaintsuan T},
  journal={Scripta Materialia},
  volume={130},
  pages={96--99},
  year={2017},
  publisher={Elsevier}
}

@article{zhang2017dislocation,
  title={Dislocation mechanisms and 3D twin architectures generate exceptional strength-ductility-toughness combination in CrCoNi medium-entropy alloy},
  author={Zhang, Zijiao and Sheng, Hongwei and Wang, Zhangjie and Gludovatz, Bernd and Zhang, Ze and George, Easo P and Yu, Qian and Mao, Scott X and Ritchie, Robert O},
  journal={Nature communications},
  volume={8},
  number={1},
  pages={14390},
  year={2017},
  publisher={Nature Publishing Group UK London}
}

@article{bacon1980anisotropic,
  title={Anisotropic continuum theory of lattice defects},
  author={Bacon, DJ and Barnett, DM and Scattergood, Ronald Otto},
  journal={Progress in Materials Science},
  volume={23},
  pages={51--262},
  year={1980},
  publisher={Elsevier}
}

@article{larsen2016robust,
doi = {10.1088/0965-0393/24/5/055007},
url = {https://doi.org/10.1088/0965-0393/24/5/055007},
year = {2016},
month = {may},
publisher = {IOP Publishing},
volume = {24},
number = {5},
pages = {055007},
author = {Larsen, Peter Mahler and Schmidt, Soren and Schiotz, Jakob},
title = {Robust structural identification via polyhedral template matching},
journal = {Modelling and Simulation in Materials Science and Engineering}
}

@article{xie2015anew,
title = {A new strain-rate-induced deformation mechanism of Cu nanowire: Transition from dislocation nucleation to phase transformation},
journal = {Acta Materialia},
volume = {85},
pages = {191-198},
year = {2015},
issn = {1359-6454},
doi = {https://doi.org/10.1016/j.actamat.2014.11.017},
url = {https://www.sciencedirect.com/science/article/pii/S1359645414008611},
author = {Hongxian Xie and Fuxing Yin and Tao Yu and Guanghong Lu and Yongguang Zhang},
}

@article{luo2019plastic,
title = {Plastic deformation induced hexagonal-close-packed nickel nano-grains},
journal = {Scripta Materialia},
volume = {168},
pages = {67-70},
year = {2019},
issn = {1359-6462},
doi = {https://doi.org/10.1016/j.scriptamat.2019.04.024},
url = {https://www.sciencedirect.com/science/article/pii/S1359646219302350},
author = {Z.P. Luo and X.K. Guo and J.X. Hou and X. Zhou and X.Y. Li and K. Lu},
}

@article{guo2021plastic,
title = {Plastic deformation induced extremely fine nano-grains in nickel},
journal = {Materials Science and Engineering: A},
volume = {802},
pages = {140664},
year = {2021},
issn = {0921-5093},
doi = {https://doi.org/10.1016/j.msea.2020.140664},
url = {https://www.sciencedirect.com/science/article/pii/S0921509320317275},
author = {X.K. Guo and Z.P. Luo and X.Y. Li and K. Lu},
}

@article{he2021stacking,
author = {He, Haiyan and Naeem, Muhammad and Zhang, Fan and Zhao, Yilu and Harjo, Stefanus and Kawasaki, Takuro and Wang, Bing and Wu, Xuelian and Lan, Si and Wu, Zhenduo and Yin, Wen and Wu, Yuan and Lu, Zhaoping and Kai, Ji-Jung and Liu, Chain-Tsuan and Wang, Xun-Li},
title = {Stacking Fault Driven Phase Transformation in CrCoNi Medium Entropy Alloy},
journal = {Nano Letters},
volume = {21},
number = {3},
pages = {1419-1426},
year = {2021},
doi = {10.1021/acs.nanolett.0c04244},
    note ={PMID: 33464087},
}

@article{li2018transformation,
title = {Transformation induced softening and plasticity in high entropy alloys},
journal = {Acta Materialia},
volume = {147},
pages = {35-41},
year = {2018},
issn = {1359-6454},
doi = {https://doi.org/10.1016/j.actamat.2018.01.002},
url = {https://www.sciencedirect.com/science/article/pii/S1359645418300338},
author = {Jia Li and Qihong Fang and Bin Liu and Yong Liu},
}

@article{pberef,
  title = {Generalized Gradient Approximation Made Simple},
  author = {Perdew, John P. and Burke, Kieron and Ernzerhof, Matthias},
  journal = {Phys. Rev. Lett.},
  volume = {77},
  issue = {18},
  pages = {3865--3868},
  numpages = {0},
  year = {1996},
  month = {Oct},
  publisher = {American Physical Society},
  doi = {10.1103/PhysRevLett.77.3865},
  url = {https://link.aps.org/doi/10.1103/PhysRevLett.77.3865}
}

@article{pawref,
  title = {Projector augmented-wave method},
  author = {Bl\"{o}chl, P. E.},
  journal = {Phys. Rev. B},
  volume = {50},
  issue = {24},
  pages = {17953--17979},
  numpages = {0},
  year = {1994},
  month = {Dec},
  publisher = {American Physical Society},
  doi = {10.1103/PhysRevB.50.17953},
  url = {https://link.aps.org/doi/10.1103/PhysRevB.50.17953}
}

@article{vasp1,
  title = {Ab initio molecular dynamics for liquid metals},
  author = {Kresse, G. and Hafner, J.},
  journal = {Phys. Rev. B},
  volume = {47},
  issue = {1},
  pages = {558--561},
  numpages = {0},
  year = {1993},
  month = {Jan},
  publisher = {American Physical Society},
  doi = {10.1103/PhysRevB.47.558},
  url = {https://link.aps.org/doi/10.1103/PhysRevB.47.558}
}

@article{vasp2,
  title = {Efficient iterative schemes for ab initio total-energy calculations using a plane-wave basis set},
  author = {Kresse, G. and Furthm\"{u}ller, J.},
  journal = {Phys. Rev. B},
  volume = {54},
  issue = {16},
  pages = {11169--11186},
  numpages = {0},
  year = {1996},
  month = {Oct},
  publisher = {American Physical Society},
  doi = {10.1103/PhysRevB.54.11169},
  url = {https://link.aps.org/doi/10.1103/PhysRevB.54.11169}
}

@article{vasp3,
title = {Efficiency of ab-initio total energy calculations for metals and semiconductors using a plane-wave basis set},
journal = {Computational Materials Science},
volume = {6},
number = {1},
pages = {15-50},
year = {1996},
issn = {0927-0256},
doi = {https://doi.org/10.1016/0927-0256(96)00008-0},
url = {https://www.sciencedirect.com/science/article/pii/0927025696000080},
author = {G. Kresse and J. Furthm\"{u}ller},
}

@article{vasp4,
  title = {Ab initio molecular-dynamics simulation of the liquid-metal--amorphous-semiconductor transition in germanium},
  author = {Kresse, G. and Hafner, J.},
  journal = {Phys. Rev. B},
  volume = {49},
  issue = {20},
  pages = {14251--14269},
  numpages = {0},
  year = {1994},
  month = {May},
  publisher = {American Physical Society},
  doi = {10.1103/PhysRevB.49.14251},
  url = {https://link.aps.org/doi/10.1103/PhysRevB.49.14251}
}

@article{vasp5,
  title = {From ultrasoft pseudopotentials to the projector augmented-wave method},
  author = {Kresse, G. and Joubert, D.},
  journal = {Phys. Rev. B},
  volume = {59},
  issue = {3},
  pages = {1758--1775},
  numpages = {0},
  year = {1999},
  month = {Jan},
  publisher = {American Physical Society},
  doi = {10.1103/PhysRevB.59.1758},
  url = {https://link.aps.org/doi/10.1103/PhysRevB.59.1758}
}

@article{meng2025akokkos,
title = {A Kokkos-accelerated Moment Tensor Potential implementation for LAMMPS},
journal = {SoftwareX},
volume = {33},
pages = {102524},
year = {2026},
issn = {2352-7110},
doi = {https://doi.org/10.1016/j.softx.2026.102524},
url = {https://www.sciencedirect.com/science/article/pii/S235271102600018X},
author = {Zijian Meng and Karim Zongo and Edmanuel Torres and Christopher Maxwell and Ryan Grant and Laurent Karim B\'{e}land},
}

@article{alber2013onrepre,
  title = {On representing chemical environments},
  author = {Bart\'{o}k, Albert P. and Kondor, Risi and Cs\'{a}nyi, G\'{a}bor},
  journal = {Phys. Rev. B},
  volume = {87},
  issue = {18},
  pages = {184115},
  numpages = {16},
  year = {2013},
  month = {May},
  publisher = {American Physical Society},
  doi = {10.1103/PhysRevB.87.184115},
  url = {https://link.aps.org/doi/10.1103/PhysRevB.87.184115}
}

@article{dscribe2,
  title={Updates to the DScribe library: New descriptors and derivatives},
  author={Laakso, Jarno and Himanen, Lauri and Homm, Henrietta and Morooka, Eiaki V and J{\"a}ger, Marc OJ and Todorovi{\'c}, Milica and Rinke, Patrick},
  journal={The Journal of Chemical Physics},
  volume={158},
  number={23},
  year={2023},
  publisher={AIP Publishing}
}

@article{wu2025revealing,
title = {Revealing the role of Al4C3 in the mechanical behavior of aluminum/graphene composites through machine learning potential-driven atomistic simulations},
journal = {Mechanics of Materials},
volume = {209},
pages = {105428},
year = {2025},
issn = {0167-6636},
doi = {https://doi.org/10.1016/j.mechmat.2025.105428},
url = {https://www.sciencedirect.com/science/article/pii/S0167663625001905},
author = {Yong-Chao Wu and Xiaoya Chang and Zhi Gen Yu and Yong-Wei Zhang and Jian-Li Shao},
}

@article{phonopy,
  author  = {Togo, Atsushi and Chaput, Laurent and Tadano, Terumasa and Tanaka, Isao},
  title   = {Implementation strategies in phonopy and phono3py},
  journal = {J. Phys. Condens. Matter},
  volume  = {35},
  number  = {35},
  pages   = {353001},
  year    = {2023},
  doi     = {10.1088/1361-648X/acd831}
}

@article{lammps,
title = {LAMMPS - a flexible simulation tool for particle-based materials modeling at the atomic, meso, and continuum scales},
journal = {Computer Physics Communications},
volume = {271},
pages = {108171},
year = {2022},
issn = {0010-4655},
doi = {https://doi.org/10.1016/j.cpc.2021.108171},
url = {https://www.sciencedirect.com/science/article/pii/S0010465521002836},
author = {Aidan P. Thompson and H. Metin Aktulga and Richard Berger and Dan S. Bolintineanu and W. Michael Brown and Paul S. Crozier and Pieter J. {in 't Veld} and Axel Kohlmeyer and Stan G. Moore and Trung Dac Nguyen and Ray Shan and Mark J. Stevens and Julien Tranchida and Christian Trott and Steven J. Plimpton},
}

@article{ovito,
doi = {10.1088/0965-0393/18/1/015012},
url = {https://doi.org/10.1088/0965-0393/18/1/015012},
year = {2009},
month = {dec},
publisher = {},
volume = {18},
number = {1},
pages = {015012},
author = {Stukowski, Alexander},
title = {Visualization and analysis of atomistic simulation data with OVITO–the Open Visualization Tool},
journal = {Modelling and Simulation in Materials Science and Engineering},
}

@article{mdapy,
title = {mdapy: A flexible and efficient analysis software for molecular dynamics simulations},
journal = {Computer Physics Communications},
volume = {290},
pages = {108764},
year = {2023},
issn = {0010-4655},
doi = {https://doi.org/10.1016/j.cpc.2023.108764},
url = {https://www.sciencedirect.com/science/article/pii/S0010465523001091},
author = {Yong-Chao Wu and Jian-Li Shao},
}

@article{gpumd,
author = {Xu, Ke and Bu, Hekai and Pan, Shuning and Lindgren, Eric and Wu, Yongchao and Wang, Yong and Liu, Jiahui and Song, Keke and Xu, Bin and Li, Yifan and Hainer, Tobias and Svensson, Lucas and Wiktor, Julia and Zhao, Rui and Huang, Hongfu and Qian, Cheng and Zhang, Shuo and Zeng, Zezhu and Zhang, Bohan and Tang, Benrui and Xiao, Yang and Yan, Zihan and Shi, Jiuyang and Liang, Zhixin and Wang, Junjie and Liang, Ting and Cao, Shuo and Wang, Yanzhou and Ying, Penghua and Xu, Nan and Chen, Chengbing and Zhang, Yuwen and Chen, Zherui and Wu, Xin and Jiang, Wenwu and Berger, Esme and Li, Yanlong and Chen, Shunda and Gabourie, Alexander J. and Dong, Haikuan and Xiong, Shiyun and Wei, Ning and Chen, Yue and Xu, Jianbin and Ding, Feng and Sun, Zhimei and Ala-Nissila, Tapio and Harju, Ari and Zheng, Jincheng and Guan, Pengfei and Erhart, Paul and Sun, Jian and Ouyang, Wengen and Su, Yanjing and Fan, Zheyong},
title = {GPUMD 4.0: A high-performance molecular dynamics package for versatile materials simulations with machine-learned potentials},
journal = {Materials Genome Engineering Advances},
volume = {3},
number = {3},
pages = {e70028},
doi = {https://doi.org/10.1002/mgea.70028},
url = {https://onlinelibrary.wiley.com/doi/abs/10.1002/mgea.70028},
year = {2025}
}

@article{atomsk,
title = {Atomsk: A tool for manipulating and converting atomic data files},
journal = {Computer Physics Communications},
volume = {197},
pages = {212-219},
year = {2015},
issn = {0010-4655},
doi = {https://doi.org/10.1016/j.cpc.2015.07.012},
url = {https://www.sciencedirect.com/science/article/pii/S0010465515002817},
author = {Pierre Hirel},
}

@article{coury2018high,
  title={High throughput discovery and design of strong multicomponent metallic solid solutions},
  author={Coury, Francisco G and Clarke, Kester D and Kiminami, Claudio S and Kaufman, Michael J and Clarke, Amy J},
  journal={Scientific reports},
  volume={8},
  number={1},
  pages={8600},
  year={2018},
  publisher={Nature Publishing Group UK London}
}

@article{esfandiarpour2025ML,
author = {Esfandiarpour, Amin and Nori, Sri Tapaswi and Bonfanti, Silvia and Alava, Mikko and Wadowski, Antoni and Huo, Wenyi and Kurpaska, Lukasz and Pecelerowicz, Michal and Wr\'{o}bel, Jan S.},
title = {Machine Learning Applied to High Entropy Alloys under Irradiation},
journal = {Advanced Engineering Materials},
volume = {27},
number = {23},
pages = {e202402280},
doi = {https://doi.org/10.1002/adem.202402280},
year = {2025}
}

@article{niu2018magnetically,
  title={Magnetically-driven phase transformation strengthening in high entropy alloys},
  author={Niu, Changning and LaRosa, Carlyn R and Miao, Jiashi and Mills, Michael J and Ghazisaeidi, Maryam},
  journal={Nature communications},
  volume={9},
  number={1},
  pages={1363},
  year={2018},
  publisher={Nature Publishing Group UK London}
}

\end{document}